\documentclass[a4paper,11pt]{article}
\pdfoutput=1 

\usepackage{jheppub} 

\usepackage[T1]{fontenc} 

\usepackage{bbm}
\usepackage{nicefrac}
\usepackage{pdflscape}
\usepackage{longtable}
\usepackage{xcolor}
\usepackage{enumerate}

\newcommand{\rep}[1]{\ensuremath\boldsymbol{#1}}

\newcommand{\Z}[1]{\ensuremath{\mathbbm{Z}_{#1}}} 
\newcommand{\SO}[1]{\ensuremath{\mathrm{SO}(#1)}}
\newcommand{\SU}[1]{\ensuremath{\mathrm{SU}(#1)}}
\newcommand{\U}[1]{\ensuremath{\mathrm{U}(#1)}}
\newcommand{\E}[1]{\ensuremath{\mathrm{E}_{#1}}}
\newcommand{\I}{\mathrm{i}}
\newcommand{\Id}{\mathbbm{1}}
\newcommand{\com}[2]{\lbrack #1, #2\rbrack}

\title{\boldmath Note on the space group selection rule for closed strings on orbifolds}

\author[a,1]{Sa\'ul~Ramos-S\'anchez,\note{Corresponding author.}}
\emailAdd{ramos@fisica.unam.mx}
\affiliation[a]{Instituto de F\'isica, Universidad Nacional Aut\'onoma de M\'exico,\\POB 20-364, Cd.Mx. 01000, M\'exico}

\author[b]{Patrick K.S. Vaudrevange}
\emailAdd{patrick.vaudrevange@tum.de}
\affiliation[b]{Physik Department T75, Technische Universit\"at M\"unchen,\\James-Franck-Stra\ss e, 85748 Garching, Germany}

\abstract{It is well-known that the space group selection rule constrains the interactions of 
closed strings on orbifolds. For some examples, this rule has been described by an effective 
Abelian symmetry that combines with a permutation symmetry to a non-Abelian flavor symmetry 
like $D_4$ or $\Delta(54)$. However, the general case of the effective Abelian symmetries was 
not yet fully understood. In this work, we formalize the computation of the Abelian symmetry 
that results from the space group selection rule by imposing two conditions only: (i) well-defined 
discrete charges and (ii) their conservation. The resulting symmetry, which we call the space group 
flavor symmetry $D_S$, is uniquely specified by the Abelianization of the space group. For all 
Abelian orbifolds with $\mathcal{N}=1$ supersymmetry we compute $D_S$ and identify new cases, for 
example, where $D_S$ contains a $\Z{2}$ dark matter-parity with charges 0 and 1 for massless and 
massive strings, respectively.}

\begin{document} 
\vspace*{-1cm}
\begin{flushright}
TUM-HEP 1170/18
\end{flushright}
\maketitle
\flushbottom

\section{Introduction}

One of the main goals of string phenomenology is to build a realistic string model that describes 
and explains the origin of the particles of the Standard Model (SM) and their interactions, as well 
as of the standard cosmology encoded in the $\Lambda$CDM model. In this sense, the compactification 
of the heterotic string on six-dimensional orbifolds~\cite{Dixon:1985jw,Dixon:1986jc} can be seen 
as a promising string scenario. It has led to very large sets of semi-realistic string models, 
which, among other features, reproduce the particle content of the 
SM~\cite{Blaszczyk:2014qoa,Blaszczyk:2015zta} or of its supersymmetric 
extensions~\cite{Lebedev:2006kn,Lebedev:2008un,Pena:2012ki,Nibbelink:2013lua,Nilles:2014owa,Carballo-Perez:2016ooy,Ramos-Sanchez:2017lmj,Olguin-Trejo:2018wpw}.

Having large sets of semi-realistic orbifold models, the next step is to improve the phenomenological 
constraints on these models. This can be achieved by a detailed study of the low-energy effective field 
theory in these constructions. First of all, this requires the identification and understanding of 
all symmetries that govern the interactions of closed-strings on orbifolds. These symmetries 
include gauge symmetries, discrete $R$-symmetries~\cite{Nilles:2013lda,Bizet:2013wha,Nilles:2017heg}, 
target-space modular symmetries~\cite{Lauer:1989ax,Lauer:1990tm,Ibanez:1992hc,Bailin:1993ri} and Abelian discrete 
symmetries that arise from the so-called space group selection rule~\cite{Hamidi:1986vh,Dixon:1986qv}. 
The latter ones are the main focus of this work: the space group selection rule sets the geometric 
conditions under which closed strings can split and join while propagating on the surface of an 
orbifold. These geometric conditions depend on the geometrical orbifold under consideration, 
which is specified by the space group $S$. 
Then, for a given space group $S$ one can rephrase the space group selection rule as an effective 
discrete symmetry which we denote by $D_S$ in the following.

The phenomenological relevance of $D_S$ has been emphasized by showing 
that they are essential ingredients of the non-Abelian flavor symmetries realized in the effective 
field theory of orbifold compactifications~\cite{Kobayashi:2006wq,Nilles:2012cy,Olguin-Trejo:2018wpw}. 
Therefore, we call the Abelian symmetry $D_S$ the {\it space group (SG) flavor symmetry}. 
Unfortunately, the nature of the SG flavor symmetry $D_S$ has not been fully understood. 
Furthermore, the computation of $D_S$ has only been restricted to the interactions of massless 
strings, even though massive strings may also play a crucial role for the phenomenology in orbifold 
compactifications, for example for CP violation due to the presence of heavy string 
modes~\cite{Nilles:2018wex}.

In this work, we aim at completing the study of the SG flavor symmetries arising from the space group 
selection rule for closed strings on orbifolds. First, in section~\ref{sec:spacegroupselectionrule} we formalize the 
computation of the SG flavor symmetry $D_S$ by imposing two physical conditions only: $D_S$ must be 
as symmetry with well-defined discrete charges for closed string states and discrete charges must be  
conserved. This fixes the SG flavor symmetry uniquely via the so-called Abelianization of the 
space group $S$. After exemplifying the computation, we present in section~\ref{sec:results} the 
SG flavor symmetries $D_S$ for all space groups $S$ of six-dimensional orbifold geometries 
with Abelian point groups and $\mathcal{N}=1$ supersymmetry as classified in ref.~\cite{Fischer:2012qj}. 
As a consistency check, we verify that all discrete anomalies of the SG flavor symmetry $D_S$ are 
universal and, hence, can be canceled by a discrete version of the universal Green-Schwarz 
mechanism~\cite{Araki:2007zza}.

Interestingly, we identify new SG flavor symmetries that act differently on massless and 
massive strings. In other words, the full string spectrum is subject to a larger symmetry group 
compared to the massless spectrum. In order to highlight possible phenomenological consequences of 
this, we present in section~\ref{sec:Z2xZ2Example51} a detailed example for a specific 
$\Z{2}\times\Z{2}$ orbifold. There, the SG flavor symmetry contains a $\Z{4}$ factor that acts as 
a $\Z{2}$ symmetry if restricted to massless strings only. As we show, such a symmetry can be used 
to define a $\Z{2}$ dark matter parity, where certain massive strings can only be produced and 
annihilated in pairs.

\section{Consequences of the space group selection rule}
\label{sec:spacegroupselectionrule}

In this section, we derive the SG flavor symmetry $D_S$ that emerges from the geometric 
restrictions on the interactions of closed strings while moving on the surface of an orbifold. 
The relevant definitions are reviewed in appendix~\ref{app:SpaceGroupsAndOrbifolds}.

\subsection{The space group selection rule}
Let us consider a coupling 
\begin{equation}\label{eq:couplingconstrelements}
|[g_1]\rangle\, |[g_2]\rangle\, \ldots\, |[g_L]\rangle
\end{equation}
of $L$ closed string states $|[g_a]\rangle$ moving on an orbifold defined by a space group $S$. 
The closed string states are characterized by their constructing elements $g_a \in S$ for 
$a=1, \ldots, L$. This coupling is allowed by the so-called space group selection 
rule~\cite{Hamidi:1986vh,Dixon:1986qv} if one can choose elements $h_a \in S$, such that
\begin{equation}\label{eqn:spacegroupselectionrule}
\prod_{a=1}^L\, h_a\, g_a\, h_a^{-1} ~=~ \Id_S\;,
\end{equation}
where $\Id_S = (\Id,0)$ is the identity element of the space group $S$. As a remark, if 
eq.~\eqref{eqn:spacegroupselectionrule} is satisfied, then the order of string states in 
eq.~\eqref{eq:couplingconstrelements} does not matter. 

Depending on the complexity of the space group $S$ under consideration, the space group selection 
rule is difficult to apply as one has to check eq.~\eqref{eqn:spacegroupselectionrule} for all 
possible choices $h_a \in S$ for $a=1,\ldots,L$. Hence, we want to identify the effective SG flavor 
symmetry $D_S$ that incorporates the space group selection rule without this ambiguity. Such 
a symmetry has to fulfill certain conditions that we now discuss.

\subsection{Conditions on the effective symmetry}
We look for a mapping, denoted by $s$, from the space group $S$ to a 
discrete group $D_S$, i.e.\ $s: S \rightarrow D_S$, with the following two properties
\begin{enumerate}[(i)]
\item $s$ is a class function:
\begin{equation}\label{eqn:classfunction}
s(h\,g\,h^{-1}) ~=~ s(g) \quad \text{for all} \quad h,g \in S\;.
\end{equation}
\item $s$ is a representation of the space group $S$: 
\begin{equation}\label{eqn:srepresentation}
s(g_1)\, s(g_2) ~=~ s(g_1\, g_2)\;,
\end{equation}
\end{enumerate}

Let us briefly discuss some immediate implications of these conditions. A necessary condition for the 
space group selection rule~\eqref{eqn:spacegroupselectionrule} to be satisfied is given by
\begin{equation}\label{eqn:effectivespacegroupselectionrule}
s(g_1) \cdots s(g_L) ~=~ s(\Id_S)\;,
\end{equation}
which follows from eq.~\eqref{eqn:spacegroupselectionrule} by applying $s$ on both sides and using 
our conditions (i) and (ii). Thus, we arrive at a much simpler version of the space group 
selection rule, where the ambiguity of choosing the elements $h_a \in S$ for $a=1,\ldots,L$ has been 
eliminated. Moreover, from a physical point of view, condition (ii) can be understood as a discrete 
charge conservation, where $s(g_a)$ corresponds to the discrete charge of a string state 
$|[g_a]\rangle$ with constructing element $g_a$, i.e.
\begin{equation}
|[g_a]\rangle ~\mapsto~ s(g_a)\, |[g_a]\rangle\;,
\end{equation}
for $a=1,\ldots,L$. Condition (i) ensures that the discrete charge corresponding to $s(g_a)$ is 
uniquely defined as a string state $|[g_a]\rangle$ with constructing element $g_a \in S$ is 
characterized by the conjugacy class $[g_a]$, see appendix~\ref{app:StringsOnOrbifolds}. 

Using the conditions (i) and (ii) of eqs.~\eqref{eqn:classfunction} 
and~\eqref{eqn:srepresentation}, it turns out that $s(g)$ (and consequently the SG flavor symmetry 
$D_S$) is fully specified for all space groups $S$. In the remainder of this paper we will compute 
our main result: the space group flavor symmetries $D_S$ for all six-dimensional orbifold 
geometries with Abelian point groups and $\mathcal{N}=1$ supersymmetry, as classified in 
ref.~\cite{Fischer:2012qj}.

As a remark, if a coupling is not invariant under $D_S$ then the space group selection rule is also 
not fulfilled. However, eq.~\eqref{eqn:effectivespacegroupselectionrule} is a necessary condition 
but not sufficient: there are cases, where a coupling is invariant under $D_S$, i.e.\ 
eq.~\eqref{eqn:effectivespacegroupselectionrule} is satisfied, but the space group selection 
rule~\eqref{eqn:spacegroupselectionrule} is not satisfied for any choice $h_a \in S$. This 
happens for example for the $\Z{6}$--II--1--1 orbifold, where the charges under $D_S$ do not depend 
on $n_1$ and $n_2$, i.e.\ on the localization of a twisted string in the first two-torus (often 
called the $G_2$ torus), see table~\ref{tab:results}. However, the space group selection rule 
applied to the first two-torus still constrains the allowed interactions~\cite{Buchmuller:2006ik}. 
Consequently, in such a case the SG flavor symmetry $D_S$ is not equivalent to the space group 
selection rule. It remains unknown whether one can modify our assumptions 
eqs.~\eqref{eqn:classfunction} and~\eqref{eqn:srepresentation} such that the resulting discrete 
symmetry is fully equivalent to the space group selection rule.

\subsection{Consequences of the conditions}
Since $s$ is a representation of the space group $S$ we 
easily see that
\begin{equation}
s(\Id_S)\, s(\Id_S) ~=~ s(\Id_S\, \Id_S) ~=~ s(\Id_S) \quad\Rightarrow\quad s(\Id_S) ~=~ \Id_{D_S}\;,
\end{equation}
where $\Id_{D_S}$ is the identity element of $D_S$. Another consequence of $s$ being a 
representation of $S$ reads
\begin{equation}\label{eqn:DSinverse}
\Id_{D_S} ~=~ s(\Id_S) ~=~ s(g^{-1}\, g) ~=~ s(g^{-1})\, s(g) \quad\Rightarrow\quad s(g^{-1}) ~=~ s(g)^{-1}\;,
\end{equation}
for all $g \in S$. In addition, eq.~\eqref{eqn:classfunction} yields
\begin{equation}\label{eqn:DSAbelian}
s(h\,g) ~=~ s(g\,h) \quad\Leftrightarrow\quad [s(h), s(g)] ~=~ 0 \;,
\end{equation}
for all $h,g \in S$. Thus, the SG flavor symmetry $D_S$ must be Abelian and the representation 
$s$ is in general not faithful. Furthermore, eqs.~\eqref{eqn:DSinverse} and~\eqref{eqn:DSAbelian} 
yield the important condition
\begin{equation}\label{eqn:SOfCommutator}
s([g,h]) ~=~ \Id_{D_S}\;,
\end{equation}
where the commutator is defined as $[g,h] = g^{-1}h^{-1}g\,h$, see appendix~\ref{app:PresentationOfS}. 
Consequently, the representation $s$ of the space group $S$ is equivalent to the Abelianization of 
the space group $S$, 
\begin{equation}
D_S ~\cong ~ S/[S,S]\;,
\end{equation}
see e.g.\ ref.~\cite{Ratcliffe:2009}.\footnote{The Abelianization of space groups was also used in 
ref.~\cite{Blaszczyk:2012} to constrain the gauge embeddings via shifts and Wilson lines, and in 
relation to Gauged Linear Sigma Models.}

Indeed, it turns out that $D_S$ is a \emph{finite} Abelian group for all space groups $S$ under 
consideration (i.e.\ in all cases the Abelian group $D_S$ does not contain any factors of $\Z{}$). 
Hence, we can represent $s(g)$ for all $g\in S$ by a complex phase, 
\begin{equation}
s(g) ~=~ \exp\left(\I\,\alpha(g)\right)\;,
\end{equation}
and set $\Id_{D_S}=1$ in the following. In addition, $D_S$ is a direct product of $\Z{M_i}$ factors 
of various orders $M_i$.

In order to obtain the discrete transformation $s(g)$ for a string state $|[g]\rangle$ with 
constructing element $g \in S$ explicitly, we first express $g$ as a product of the generators of 
$S$, see appendix~\ref{app:SpaceGroup}. In detail, in the case of a $\Z{M}\times\Z{N}$ point group 
the generators of $S$ read
\begin{equation}
T_i ~=~ (\Id, e_i) \;,\, g_\theta ~=~ (\theta, \lambda_\theta) \quad\text{and}\quad g_\omega ~=~ (\omega, \lambda_\omega)\;,
\end{equation}
such that we can decompose any space group element $g \in S$ as
\begin{equation}\label{eqn:GeneralSpaceGroupElement}
g ~=~ \left(T_1\right)^{n_1}\,\ldots \left(T_D\right)^{n_D}\, \left(g_\theta\right)^k\, \left(g_\omega\right)^\ell\;,
\end{equation}
where $k=0,\ldots,M-1$, $\ell=0,\ldots,N-1$ and $n_i \in\Z{}$ for $i=1,\ldots,D$. Then, using the 
representation property eq.~\eqref{eqn:srepresentation}, we obtain
\begin{equation}\label{eqn:RepOfGeneralSpaceGroupElement}
s(g) ~=~ \left(s(T_1)\right)^{n_1}\,\ldots \left(s(T_D)\right)^{n_D}\, \left(s(g_\theta)\right)^k\, \left(s(g_\omega)\right)^\ell\;.
\end{equation}
Thus, we can easily compute the discrete transformation $s(g)$ of a string state $|[g]\rangle$ 
once we know the discrete transformations of the generators of the space group 
\begin{equation}
s(T_i) \;,\; s(g_\theta) \quad\text{and}\quad s(g_\omega)\;.
\end{equation}
Hence, we have reduced our problem to the task of identifying these building-blocks of the general 
discrete transformation $s(g)$.

\subsection{Space group selection rule and the Abelianization of the space group}

As noted before, our conditions eqs.~\eqref{eqn:classfunction} and~\eqref{eqn:srepresentation} that 
define the transformation $s(g)$ of a closed string state $|[g]\rangle$ with constructing element 
$g \in S$ correspond to the Abelianization of the space group $S$. In this section we show that the 
Abelianization of $S$ can be computed using the so-called presentation of the space group $S$. To 
do so, each generator $g$ of $S$ is replaced by $s(g)$ and each relation in the presentation of $S$ 
is replaced by $s(\text{relation})$ using that $s([g,h])=1$, i.e.\
\begin{subequations}
\begin{eqnarray}
g_\theta ~=~ (\theta, \lambda_\theta) & ~\longmapsto~ & s(g_\theta) \;,\\
g_\omega ~=~ (\omega, \lambda_\omega) & ~\longmapsto~ & s(g_\omega) \;,\\
T_i      ~=~ (\Id,    e_i)            & ~\longmapsto~ & s(T_i) \;, \quad i=1,\ldots,D\;,
\end{eqnarray}
\end{subequations}
and from the presentation
\begin{equation}
S ~=~ \langle g_\theta,\, g_\omega,\, T_1,\, \ldots,\, T_D ~|~ g_\theta^{M}\, \left(T_1\right)^{-a_{(\theta,1)}}\, \ldots\, \left(T_D\right)^{-a_{(\theta,D)}},\, \ldots\,,\, \text{all relations} \rangle\;,
\end{equation}
see appendix~\ref{app:PresentationOfS}, we obtain via the map $S \mapsto D_S$ the presentation of 
$D_S$ as
\begin{eqnarray}\label{eqn:PresentationOfDS}
D_S ~=  && \langle s(g_\theta),\, s(g_\omega),\, s(T_1),\, \ldots,\, s(T_D) ~|~ \nonumber\\
        && \;\;s(g_\theta)^{M}\, \left(s(T_1)\right)^{-a_{(\theta,1)}}\, \ldots\, \left(s(T_D)\right)^{-a_{(\theta,D)}},\,\ldots\,,\, s(\text{all relations}) \rangle\;.
\end{eqnarray}
In most cases, some generators $s(g)$ in the presentation of $D_S$ are no longer independent 
compared to their preimages $g\in S$. Thus, one has to solve the relations in $D_S$ such that only 
the independent generators remain. 

Furthermore, for all space groups under consideration all remaining relations in $D_S$ can be 
solved explicitly such that one can identify the SG flavor symmetry $D_S$ as the direct product of 
cyclic groups, i.e.
\begin{equation}
D_S ~\cong ~ \Z{M_1} \times \Z{M_2} \times \ldots\;.
\end{equation}
In the following we present two approaches to how this computation can be performed in detail.

\subsection{Direct computation}

Among the relations in the presentation of $D_S$ in eq.~\eqref{eqn:PresentationOfDS} there are 
two sets of relations that are of special interest: eqs.~\eqref{eqn:RelationThetaTranslation} 
and~\eqref{eqn:RelationOmegaTranslation} (or equivalently eq.~\eqref{eqn:conjugatetranslation}) are 
related to the charges of translations $s(T_i)$, while 
eqs.~\eqref{eqn:relationOrderM} and~\eqref{eqn:relationOrderN} are related to the charges of 
rotations $s(g_\theta)$ and $s(g_\omega)$. Using these relations one can easily compute the orders of 
$s(g_\theta)$, $s(g_\omega)$ and $s(T_i)$, as we do next.

\paragraph{\boldmath The charges of translations $s(T_i)$.\unboldmath}
We embed eq.~\eqref{eqn:conjugatetranslation} into $D_S$ and obtain\footnote{For a space group 
element $g = (\rho, \lambda) \in S$ we write $s(g)=s(\rho,\lambda)$ instead of $s(g)=s((\rho,\lambda))$.}
\begin{equation}\label{eqn:rotidentificationofs}
s(\Id,e_i) ~=~ s(\Id,\rho\,e_i)\;,
\end{equation}
for all $\rho \in P$. In other words, two vectors $e_i$ and $\rho\,e_i$ that are rotated to 
each other by a point group element $\rho$ give rise to the same element of $D_S$. 

Now, we assume that there is a point group element $\rho \in P$ and a smallest integer $M_i$ such 
that
\begin{equation}\label{eqn:orderoftranslation}
e_i + \rho\, e_i + \ldots + \rho^{M_i - 1}\, e_i ~=~ 0\;.
\end{equation}
We can translate this equation into space group elements of pure translations and apply $s$. This 
yields
\begin{equation}
\underbrace{s(\Id,e_i)\, s(\Id,\rho\, e_i) \ldots s(\Id,\rho^{M_i - 1}\, e_i)}_{M_i \text{ factors}} ~=~ 1\;.
\end{equation}
Then, using eq.~\eqref{eqn:rotidentificationofs} we get
\begin{equation}
s(\Id,e_i)^{M_i} ~=~ 1\;.
\end{equation}
Consequently, $s(T_i)=s(\Id,e_i)$ generates a finite Abelian group of order $M_i$, being $\Z{M_i}$, and we can 
express $s(T_i)$ as a complex phase
\begin{equation}
s(T_i) ~=~ \exp\left(\frac{2\pi\I}{M_i} \beta_i \right)\;,
\end{equation}
for $\beta_i \in \Z{}$. 

\paragraph{\boldmath The charges of rotations $s(g_\theta)$ and $s(g_\omega)$.\unboldmath}
For example, consider the generator $\theta \in P$ of order $M$ in the case without 
roto-translations, i.e.\ $g_\theta =(\theta,0)$. Then, we obtain the following identities
\begin{equation}
s(\theta,0)^M ~=~ s(\theta^M,0) ~=~ s(\Id_S) ~=~ 1\;.
\end{equation}
Consequently, $s(g_\theta)=s(\theta,0)$ generates a finite Abelian group of order $M$, being 
$\Z{M}$, and we can express $s(g_\theta)$ as a complex phase
\begin{equation}
s(g_\theta) ~=~ \exp\left(\frac{2\pi\I}{M} \alpha_\theta \right)\;,
\end{equation}
for $\alpha_\theta \in \Z{}$. Analogously, using $\omega^N = \Id$ we find
\begin{equation}
s(g_\omega) ~=~ \exp\left(\frac{2\pi\I}{N} \alpha_\omega \right)\;,
\end{equation}
for $\alpha_\omega \in \Z{}$.

\paragraph{\boldmath Combination of translations and rotations.\unboldmath}
As we have seen in eq.~\eqref{eqn:RepOfGeneralSpaceGroupElement} the representation $s(g)$ of a general 
space group element eq.~\eqref{eqn:GeneralSpaceGroupElement} is given by
\begin{equation}
s(g) ~=~ \left(s(T_1)\right)^{n_1}\,\ldots \left(s(T_D)\right)^{n_D}\, \left(s(g_\theta)\right)^k\, \left(s(g_\omega)\right)^\ell\;,
\end{equation}
where one might be tempted to simply replace the previous results for the representations
of the translations, $s(T_i)$, and roto-translations, $s(g_\theta),s(g_\omega)$. However, 
in general, the presentation of $D_S$ eq.~\eqref{eqn:PresentationOfDS} establishes non-trivial 
relations among different elements, that can alter the result.

In fact, only in the simplest case without roto-translations and without non-trivial 
relations between translations and twists, this na\"ive expectation holds.
In this case, a general space group element simplifies to $g=(\theta^k\, \omega^\ell, n_i\, e_i) \in S$ 
and its representation is given by
\begin{equation}
s \; : \; (\theta^k\, \omega^\ell, n_i\, e_i) ~\rightarrow~ \exp\left(2\pi\I \left(\frac{k}{M} \alpha_\theta + \frac{\ell}{N} \alpha_\omega + \sum_{i=1}^D\,\frac{n_i}{M_i} \beta_j\right)\right)\;,
\end{equation}
where $k$, $\ell$ and $n_i$ for $i=1,\ldots, D$ are discrete charges. They are conserved in an 
allowed coupling, i.e.
\begin{equation}
\sum_{a=1}^L\, k^{(a)}    ~=~ 0 \text{ mod } M\;, \quad \sum_{a=1}^L\, \ell^{(a)} ~=~ 0 \text{ mod } N\;,\quad \sum_{a=1}^L\, n_i^{(a)}  ~=~ 0 \text{ mod } M_i\;,
\end{equation}
where $k^{(a)}$, $\ell^{(a)}$ and $n_i^{(a)}$ define the $a$-th constructing element $g_a \in S$
in the coupling~\eqref{eq:couplingconstrelements}.
As a remark, those symmetries that constrain the twisted sectors $k^{(a)}$ (and $\ell^{(a)}$) are 
conventionally referred to as point group selection rule (PG).

In the more general case, $\alpha_\theta$, $\alpha_\omega$, $\beta_i$ and $\beta_j$, with $i\neq j$, 
turn out to be connected by the relations in the presentation of $D_S$. These relations as well 
as their implications shall be discussed in our examples, in section~\ref{sec:Examples}.

\subsection{Space group selection rule and remnant discrete symmetries from spontaneous symmetry breaking}
\label{sec:VEVMethod}

As before, we begin with a space group $S$ generated by $D+2$ generators
\begin{equation}
T_i ~=~ (\Id, e_i)\;,\; i=1,\ldots,D\;,\; g_\theta ~=~ (\theta, \lambda_\theta) \quad\text{and}\quad g_\omega ~=~ (\omega, \lambda_\omega)
\end{equation}
subject to $K$ relations as listed in the presentation of $S$. Then, the SG flavor symmetry $D_S$ 
obtained from the space group selection rule is a subgroup of $\U{1}^{D+2}$. It can be computed as 
follows: each relation $\alpha=1,\ldots,K$ of the presentation of $S$ is mapped to a relation in 
$D_S$ and can be written in the form
\begin{equation}\label{eqn:SSBRelation}
s(T_1)^{q^\alpha_1}\,\ldots\, s(T_D)^{q^\alpha_D}\, s(g_\theta)^{q^\alpha_{D+1}}\, s(g_\omega)^{q^\alpha_{D+2}} ~=~ 1\;,
\end{equation}
see eq.~\eqref{eqn:SpaceGroupRelations} in appendix~\ref{app:PresentationOfS} using $s([g,h])=1$. 
To each relation~\eqref{eqn:SSBRelation} one can associate an auxiliary field $\phi_\alpha(x)$ with 
integer $\U{1}^{D+2}$ charges given by the exponents in eq.~\eqref{eqn:SSBRelation}, i.e.\ the 
charges of $\phi_\alpha(x)$ read
\begin{equation}
(q^\alpha_1,\ldots, q^\alpha_D, q^\alpha_{D+1}, q^\alpha_{D+2})\;.
\end{equation}
These auxiliary fields have no physical interpretation but are only used to describe the symmetry 
breaking associated to the relations~\eqref{eqn:SSBRelation}. Then, under a general $\U{1}^{D+2}$ 
transformation the auxiliary field $\phi_\alpha(x)$ picks up a phase, i.e.\
\begin{equation}
\phi_\alpha(x) ~\mapsto~ \exp\left(\I \sum_{j=1}^{D+2} \xi_j\, q^\alpha_j \right)\,\phi_\alpha(x)\;,
\end{equation}
where $\xi_j \in \mathbbm{R}$, $j=1,\ldots,D+2$, denote the $\U{1}$ transformation parameters. 
Now, we turn on the vacuum expectation values (VEVs) of the auxiliary fields 
$\langle\phi_\alpha\rangle\neq 0$ for all $\alpha=1,\ldots,K$. Consequently, the unbroken remnant 
symmetry is given by the solutions $\xi_j$ of
\begin{equation}
\exp\left(\I \sum_{j=1}^{D+2} \xi_j\, q^\alpha_j \right)\,\langle\phi_\alpha\rangle ~=~ \langle \phi_\alpha\rangle 
\end{equation}
for all auxiliary fields $\alpha=1,\ldots,K$. Identifying
\begin{equation}
s(T_i) ~=~ \exp\left(\I\,\xi_i\right) \;,\; s(g_\theta) ~=~ \exp\left(\I\,\xi_{D+1}\right) \quad\text{and}\quad s(g_\omega) ~=~ \exp\left(\I\,\xi_{D+2}\right)\;,
\end{equation}
we realize that this spontaneous $\U{1}$ symmetry breaking exactly corresponds to solving 
the relations~\eqref{eqn:SSBRelation}.

This method to compute the SG flavor symmetry as a remnant discrete symmetry via spontaneous 
symmetry breaking of $\U{1}^{D+2}$ can be automatized easily using for example the mathematica 
package ``DiscreteBreaking'' developed in ref.~\cite{Petersen:2009ip}.

\subsection[Examples]{Examples}
\label{sec:Examples}

In this section we present three examples how to compute the SG flavor symmetries explicitly: 
first, we consider a $\Z{3}$ toy example in $D=2$ dimensions and then two $\Z{2}\times\Z{2}$ examples 
in $D=6$ dimensions, one with roto-translations and the other with a freely-acting shift.

\subsubsection[$\Z{3}$ space group in $D=2$ dimensions]{\boldmath $\Z{3}$ space group in $D=2$ dimensions\unboldmath}

We define the two-dimensional $\Z{3}$ space group $S$ by the generators
\begin{equation}
T_1 ~=~ (\Id, e_1) \;, \quad T_2 ~=~ (\Id, e_2) \quad\text{and}\quad g_\theta ~=~ (\theta, 0)\;,
\end{equation}
where the basis vectors $e_1$ and $e_2$ enclose an angle of $120^\circ$ and have equal length. 
Furthermore, $\theta$ is a counter-clockwise rotation by $120^\circ$ such that
\begin{equation}
\theta\, e ~=~ e\, \hat\theta \quad\text{where}\quad \hat\theta ~=~ \left(\begin{array}{cc} 0 & -1\\ 1 & -1\end{array}\right)\;\,.
\end{equation}
This space group can be defined alternatively by the following abstract presentation, based on the 
three generators $g_\theta$, $T_1$ and $T_2$ subject to four relations, i.e.
\begin{equation}
S ~=~ \langle g_\theta,\, T_1,\, T_2 ~|~ g_\theta^3\;, \com{T_1}{T_2}\;, \com{g_\theta}{T_1}\,T_1^{-2}\,T_2^{-1}\;, \com{g_\theta}{T_2}\,T_1\,T_2^{-1} \rangle\;,
\end{equation}
see appendix~\ref{app:PresentationOfS}.

To compute the SG flavor symmetry $D_S$ we map the three generators of $S$ to $s(g_\theta)$, 
$s(T_1)$ and $s(T_2)$. Then, the presentation of $D_S$ reads
\begin{eqnarray}
D_S ~= && \langle s(g_\theta),\, s(T_1),\, s(T_2) ~|~ s(g_\theta)^3,\, \underbrace{s(\com{T_1}{T_2})}_{=1},\, \underbrace{s(\com{g_\theta}{T_1})}_{=1}\,s(T_1)^{-2}\,s(T_2)^{-1},\, \nonumber\\
       &&\;\;\underbrace{s(\com{g_\theta}{T_2})}_{=1}\,s(T_1)\,s(T_2)^{-1} \rangle\;.
\end{eqnarray}
Next, we omit the trivial relation $s(\com{T_1}{T_2})=1$ and obtain
\begin{equation}\label{eqn:PresentationOfDSForZ3}
D_S ~=~ \langle s(g_\theta),\, s(T_1),\, s(T_2) ~|~ s(g_\theta)^3,\, s(T_1)^{-2}\,s(T_2)^{-1},\, s(T_1)\,s(T_2)^{-1} \rangle\;.
\end{equation}
In the final step, we observe that only two of the three generators of $D_S$ are independent. For 
example, we can use the last relation $s(T_1)\,s(T_2)^{-1}=1$ in eq.~\eqref{eqn:PresentationOfDSForZ3} 
to replace $s(T_2)$ by $s(T_1)$ using $s(T_2)=s(T_1)$. Thus, we get
\begin{equation}\label{eqn:FinalPresentationOfDSForZ3}
D_S ~=~ \langle s(g_\theta),\, s(T_1) ~|~ s(g_\theta)^3,\, s(T_1)^3 \rangle ~\cong~ \Z{3}\times\Z{3}\;,
\end{equation}
and a general string state $|[g]\rangle$ transforms as
\begin{eqnarray}
|[g]\rangle & ~\longmapsto~ & s(T_1)^{n_1}\,s(T_2)^{n_2}\,s(g_\theta)^{k}\,|[g]\rangle ~=~ s(T_1)^{n_1+n_2}\,s(g_\theta)^{k}\,|[g]\rangle \\
            & =             & \exp\left(\frac{2\pi\I}{3} \alpha_{1}\,(n_1+n_2)\right)\,\exp\left(\frac{2\pi\I}{3} \alpha_\theta\,k\right)\,|[g]\rangle \;,
\end{eqnarray}
for $\alpha_1$, $\alpha_\theta \in \{0,1,2\}$ using $s(g_\theta)^3 = s(T_1)^3 =1$. Then, a coupling 
$|[g_1]\rangle\,\ldots\,|[g_L]\rangle$ of string states $|[g_a]\rangle$ with constructing elements 
$g_a=(\theta^{k^{(a)}}, n_1^{(a)} e_1 + n_2^{(a)} e_2)$ is allowed by the SG flavor symmetry 
$D_S \cong \Z{3}\times\Z{3}$ if
\begin{equation}\label{eqn:Z3ExamplePG}
\sum_{a=1}^L k^{(a)}                            ~=~ 0 \;\text{ mod }\; 3 \quad\text{and}\quad
\sum_{a=1}^L \left(n_1^{(a)} + n_2^{(a)}\right) ~=~ 0 \;\text{ mod }\; 3 \;,
\end{equation}
where the first equation in eq.~\eqref{eqn:Z3ExamplePG} is called point group selection rule (PG).

In summary, the space group selection rule of the two-dimensional $\Z{3}$ space group yields a 
$\Z{3}\times\Z{3}$ SG flavor symmetry with discrete charges $k$ and $n_1+n_2$, 
respectively~\cite{Hamidi:1986vh}.

\subsubsection[Space group with freely-acting shift: $\Z{2}\times\Z{2}$--5--1]{\boldmath Space group with freely-acting shift: $\Z{2}\times\Z{2}$--5--1\unboldmath}
\label{sec:Z2xZ2Example51}

Consider the so-called Blaszczyk-geometry $\Z{2}\times\Z{2}$--5--1~\cite{Fischer:2012qj} (in 
ref.~\cite{Donagi:2008xy} it is labeled as 1--1, see also~\cite{Forste:2006wq}, 
and~\cite{Blaszczyk:2009in} for an MSSM-like orbifold model based on this geometry using a 
different convention). The twists in the lattice basis (see eq.~\eqref{eqn:TwistInLatticeBasis}) read
\begin{equation}
\hat\theta ~=~ \left(
\begin{array}{cccccc}
  0 &  1 & -1 &  0 &  0 &  0\\ 
  1 &  0 & -1 &  0 &  0 &  0\\
  0 &  0 & -1 &  0 &  0 &  0\\
  0 &  0 &  0 &  1 &  0 &  0\\
  0 &  0 &  0 &  0 & -1 &  0\\
  0 &  0 &  0 &  0 &  0 & -1\\
\end{array}
\right)
\quad\text{and}\quad
\hat\omega ~=~ \left(
\begin{array}{cccccc}
  0 & -1 &  1 &  0 &  0 &  0\\
  0 & -1 &  0 &  0 &  0 &  0\\
  1 & -1 &  0 &  0 &  0 &  0\\
  0 &  0 &  0 & -1 &  0 &  0\\
  0 &  0 &  0 &  0 &  1 &  0\\
  0 &  0 &  0 &  0 &  0 & -1\\
\end{array}
\right)\;.
\end{equation}
The space group $S$ is generated by six translations $T_i = (\Id, e_i)$ and two rotations
\begin{equation}
g_\theta ~=~ \left(\theta, 0\right) \quad\text{and}\quad g_\omega ~=~ \left(\omega, 0\right)\;.
\end{equation}
By writing down the presentation of this space group $S$ we identify the relations 
\begin{subequations}\label{eqn:Z2xZ251Relations}
\begin{eqnarray}
s(T_1)   & = & s(T_2) ~=~ s(T_3) \quad\text{where}\quad s(T_1)^2\,s(T_2)\,s(T_3) ~=~ 1\;,\label{eqn:Z2xZ251Relation1}\\
s(T_4)^2 & = & s(T_5)^2 ~=~ s(T_6)^2 ~=~ s(g_\theta)^2 ~=~ s(g_\omega)^2 ~=~ 1\;,
\end{eqnarray}
\end{subequations}
see eq.~\eqref{eqn:SpaceGroupRelations}. Consequently, the most general representation $s(g)$ is 
given by
\begin{subequations}
\begin{eqnarray}
s(g) & = & s(T_1)^{n_1}\, s(T_2)^{n_2}\, s(T_3)^{n_3}\, s(T_4)^{n_4}\, s(T_5)^{n_5}\, s(T_6)^{n_6}\, s(g_\theta)^k\, s(g_\omega)^\ell \\
     & = & s(T_1)^{n_1+n_2+n_3}\, s(T_4)^{n_4}\, s(T_5)^{n_5}\, s(T_6)^{n_6}\, s(g_\theta)^k\, s(g_\omega)^\ell\;,
\end{eqnarray}
\end{subequations}
where we have used $s(T_1) = s(T_2) = s(T_3)$. The order of the basic transformations 
$s(g_\theta)$, $s(g_\omega)$, $s(T_4)$, $s(T_5)$ and $s(T_6)$ is 2. Furthermore, from 
eq.~\eqref{eqn:Z2xZ251Relation1} we obtain
\begin{equation}
s(T_1)^2\,s(T_2)\,s(T_3) ~=~ s(T_1)^4 ~=~ 1\;,
\end{equation}
and $s(T_1)$ generates a $\Z{4}$ factor. Consequently, all transformations $s(g_\theta)$, 
$s(g_\omega)$, $s(T_1)$, $s(T_4)$, $s(T_5)$ and $s(T_6)$ are independent. Hence, the SG flavor 
symmetry $D_S$ obtained from the space group selection rule is given by
\begin{equation}\label{eqn:Z2xZ251Symmetry}
\left(\Z{2} \times \Z{2}\right)^\text{PG} \times \Z{4} \times \Z{2} \times \Z{2} \times \Z{2}\;,
\end{equation}
with discrete charges
\begin{equation}
k \;\;,\;\; \ell  \;\;,\;\; n_1+n_2+n_3 \;\;,\;\; n_4 \;\;,\;\; n_5 \quad\text{and}\quad n_6\;,
\end{equation}
respectively, and we have verified this result additionally using the VEV-method of section~\ref{sec:VEVMethod}.

\paragraph{\boldmath $\Z{2}$ dark matter-parity.\unboldmath} By considering the fixed points of 
this $\Z{2}\times\Z{2}$ orbifold, one can check that all massless strings carry even $\Z{4}$ charges, 
i.e.
\begin{equation}
n_1+n_2+n_3 ~\in~ \{0, 2\}\;.
\end{equation}
Hence, the allowed interactions of massless strings are not constrained by a $\Z{4}$ factor in 
eq.~\eqref{eqn:Z2xZ251Symmetry}, but only by $\Z{2}$. However, there are massive strings with odd 
discrete charges $n_1+n_2+n_3\in\{1,3\}$ under $\Z{4}$, for example a winded string with 
constructing element $(\Id,e_1)$. Consequently, massive strings with odd $\Z{4}$ charges can only 
be produced and annihilated in pairs. Thus, the lightest massive string with odd 
$\Z{4}$ charge could serve as a dark matter candidate, which is stable because there is a $\Z{2}$ 
dark matter-parity with
\begin{subequations}
\begin{eqnarray}
|\text{matter}\rangle      & ~\longmapsto~ & + |\text{matter}\rangle\;, \\
|\text{dark matter}\rangle & ~\longmapsto~ & - |\text{dark matter}\rangle\;,
\end{eqnarray}
\end{subequations}
and the mass of the dark matter particle depends on the compactification radii. This fact is common 
to many SG flavor symmetries in table~\ref{tab:results} and might be relevant for the 
observed dark matter content of the universe and also for its cosmological evolution.

\subsubsection[Space group with roto-translation: $\Z{2}\times\Z{2}$--2--5]{\boldmath Space group with roto-translation: $\Z{2}\times\Z{2}$--2--5\unboldmath}
\label{sec:Z2xZ2-2-4}

Consider the $\Z{2}\times\Z{2}$--2--5 orbifold from the classification in 
ref.~\cite{Fischer:2012qj} (it corresponds to the $\Z{2}\times\Z{2}$ orbifold labeled 1--9 in 
ref.~\cite{Donagi:2008xy}). The twists in the lattice basis (see eq.~\eqref{eqn:TwistInLatticeBasis}) are 
given by
\begin{equation}
\hat\theta ~=~ \left(
\begin{array}{cccccc}
 0 & 1 & 0 & 0 & 0 & 0 \\
 1 & 0 & 0 & 0 & 0 & 0 \\
 0 & 0 & 1 & 0 & 0 & 0 \\
 0 & 0 & 0 & -1 & 0 & 0 \\
 0 & 0 & 0 & 0 & -1 & 0 \\
 0 & 0 & 0 & 0 & 0 & -1 \\
\end{array}
\right)
\quad\text{and}\quad
\hat\omega ~=~ \left(
\begin{array}{cccccc}
 -1 & 0 & 0 & 0 & 0 & 0 \\
 0 & -1 & 0 & 0 & 0 & 0 \\
 0 & 0 & -1 & 0 & 0 & 0 \\
 0 & 0 & 0 & 1 & 0 & 0 \\
 0 & 0 & 0 & 0 & 1 & 0 \\
 0 & 0 & 0 & 0 & 0 & -1 \\
\end{array}
\right)\;.
\end{equation}
The space group $S$ is generated by six translations $T_i = (\Id, e_i)$ and two roto-translations
\begin{equation}
g_\theta ~=~ \left(\theta, \frac{1}{2}e_3\right) \quad\text{and}\quad g_\omega ~=~ \left(\omega, \frac{1}{2}e_5\right)\;.
\end{equation}
By writing down the presentation of $S$ we identify the relations 
(see eq.~\eqref{eqn:SpaceGroupRelations})
\begin{subequations}\label{eqn:Z2xZ225Relations}
\begin{eqnarray}
s(T_1)   & = & s(T_2) \;\;,\;\;   s(T_3) ~=~ s(T_5) ~=~ s(g_\theta)^2 ~=~ s(g_\omega)^2 \quad\text{and}\quad \label{eqn:Z2xZ225Relation1}\\
s(T_i)^2 & = & s(g_\theta)^4 ~=~ s(g_\omega)^4 ~=~ 1 \quad\text{for}\quad i = 1,\ldots,6\;.
\end{eqnarray}
\end{subequations}
Consequently, the most general charge $s(g)$ is given by
\begin{subequations}
\begin{eqnarray}
s(g) & = & s(T_1)^{n_1}\, s(T_2)^{n_2}\, s(T_3)^{n_3}\, s(T_4)^{n_4}\, s(T_5)^{n_5}\, s(T_6)^{n_6}\, s(g_\theta)^k\, s(g_\omega)^\ell \\
     & = & s(T_1)^{n_1+n_2}\, s(T_4)^{n_4}\, s(T_6)^{n_6} \, s(g_\theta)^{k+2(n_3+n_5)}\, s(g_\omega)^\ell\;,\label{eqn:Z2xZ225Charge}
\end{eqnarray}
\end{subequations}
where we have used $s(T_1) = s(T_2)$ and $s(T_3) = s(T_5) = s(g_\theta)^2$. Next, we analyze the consequences of 
$s(g_\theta)^2 = s(g_\omega)^2$ from eq.~\eqref{eqn:Z2xZ225Relation1}, i.e.\ we make the ansatz
\begin{equation}
s(g_\theta) ~=~ \exp\left(\frac{2\pi\I}{4}\,\alpha_\theta\right) \quad\text{and}\quad s(g_\omega) ~=~ \exp\left(\frac{2\pi\I}{4}\,\alpha_\omega\right) \;.
\end{equation}
Then, $s(g_\theta)^2 = s(g_\omega)^2$ yields
\begin{equation}
\exp\left(\frac{2\pi\I}{2}\,\alpha_\theta\right) ~=~ \exp\left(\frac{2\pi\I}{2}\,\alpha_\omega\right) \quad\Leftrightarrow\quad \alpha_\omega ~=~ \alpha_\theta + 2\,x\;,
\end{equation}
for some $x \in\Z{}$. Thus,
\begin{equation}\label{eqn:Z2xZ225ChargeOmegaX}
s(g_\omega) ~=~ \exp\left(\frac{2\pi\I}{4}\,\alpha_\omega\right) ~=~ \exp\left(\frac{2\pi\I}{4}\,\alpha_\theta\right)\, \exp\left(\frac{2\pi\I}{2}\,x\right) ~=~ s(g_\theta)\, s_x\;.
\end{equation}
where $s(g_\theta)$ and $s_x$ are now independent and of order 4 and 2, respectively. Using 
eq.~\eqref{eqn:Z2xZ225ChargeOmegaX} in eq.~\eqref{eqn:Z2xZ225Charge} we obtain
\begin{equation}
s(g) ~=~ s(g_\theta)^{k+\ell+2(n_3+n_5)}\, s_x^\ell\, s(T_1)^{n_1+n_2}\, s(T_4)^{n_4}\, s(T_6)^{n_6}\;,
\end{equation}
where the orders of $s(g_\theta)$, $s_x$, $s(T_1)$, $s(T_4)$ and $s(T_6)$ are 4, 2, 2, 2 and 2, respectively. 
Now, we have solved all relations~\eqref{eqn:Z2xZ225Relations} and, consequently, all 
transformations $s(g_\theta)$, $s_x$, $s(T_1)$, $s(T_4)$ and $s(T_6)$ are independent. Hence, the space group 
selection rule results in a SG flavor symmetry
\begin{equation}
\left(\Z{4} \times \Z{2}\right)^\text{PG} \times \Z{2} \times \Z{2} \times \Z{2}\;,
\end{equation}
with discrete charges
\begin{equation}
k+\ell+2(n_3+n_5) \;\;,\;\; \ell  \;\;,\;\; n_1+n_2 \;\;,\;\; n_4 \quad\text{and}\quad n_6\;,
\end{equation}
respectively. This result has also been verified using the VEV-method of section~\ref{sec:VEVMethod}.

Naively, one would expect a $\Z{2}\times\Z{2}$ point group selection rule with charges $k$ and 
$\ell$ if the point group is $\Z{2}\times\Z{2}$. However, we have seen that the point group 
selection rule yields $\Z{4}\times\Z{2}$ with charges $k+\ell+2(n_3+n_5)$ and $\ell$. If one 
considers the $\Z{2}$ subgroup of the $\Z{4}$ factor, one identifies the corresponding $\Z{2}$ 
charges as $k+\ell$. Thus, the naive $\Z{2}\times\Z{2}$ point group selection rule is a subgroup 
of the full $\Z{4}\times\Z{2}$ point group selection rule.

\section{Results}
\label{sec:results}

In this section we present the main result of this work: we compute the Abelian SG flavor 
symmetries $D_S$ for all 138 space groups $S$ with Abelian point group and $\mathcal{N}$=1 
supersymmetry. To do so, we adopt the convention of ref.~\cite{Fischer:2012qj} as specified in the 
geometry files for the \texttt{orbifolder}~\cite{Nilles:2011aj} (see ancillary files in arXiv.org). 
The results are listed in table~\ref{tab:results}.

\begin{longtable}{|c|c|l|l|}
\caption{\label{tab:results}Space group flavor symmetries obtained from the space group selection rule for all space groups with Abelian point group and $\mathcal{N}$=1 supersymmetry~\cite{Fischer:2012qj}.}
\vspace{-0.5cm}
\\
\hline
$\mathbbm{Q}$-class & $\mathbbm{Z}$- and & SG flavor          & discrete                          \\
twist vector        & affine class       & symmetry $D_S$     & charge                            \\
\hline
\hline
\endfirsthead
\hline
$\mathbbm{Q}$-class & $\mathbbm{Z}$- and & SG flavor          & discrete                          \\
twist vector        & affine class       & symmetry $D_S$     & charge                            \\
\hline
\hline
\endhead
\hline
\multicolumn{4}{r}{continued...}\\
\endfoot
\endlastfoot
$\Z{3}$              &1--1     & $\Z{3}^{(\text{PG})}$                          & $k$ \\
$\left(0,\frac{1}{3},\frac{1}{3},-\frac{2}{3}\right)$
                     &         & $\left(\Z{3}\right)^3$                         & $(n_1 + n_2, n_3 + n_4, n_5 + n_6)$ \\
\hline
$\Z{4}$              &1--1     & $\Z{4}^{(\text{PG})}$                          & $k$ \\
$\left(0,\frac{1}{4},\frac{1}{4},-\frac{1}{2}\right)$
                     &         & $\left(\Z{2}\right)^4$                         & $(n_1 + n_2, n_3 + n_4, n_5, n_6)$ \\
\cline{2-4}
                     &2--1     & $\Z{4}^{(\text{PG})}$                          & $k$ \\
                     &         & $\Z{4} \times \left(\Z{2}\right)^2$            & $(n_3 + n_4 + n_5, n_1 + n_2, n_6)$ \\
\cline{2-4}
                     &3--1     & $\Z{4}^{(\text{PG})}$                          & $k$ \\
                     &         & $\left(\Z{4}\right)^2$                         & $(n_1 + n_2 + n_3, n_4 + n_5 + n_6)$ \\
\hline
$\Z{6}$--I           &1--1     & $\Z{6}^{(\text{PG})}$                          & $k$ \\
$\left(0,\frac{1}{6},\frac{1}{6},-\frac{1}{3}\right)$
                     &         & $\Z{3}$                                        & $n_5 + n_6$ \\
\cline{2-4}
                     &2--1     & $\Z{6}^{(\text{PG})}$                          & $k$ \\
                     &         & $\Z{3}$                                        & $n_3 + n_4 + n_5 + n_6$ \\
\hline
$\Z{6}$--II          &1--1     & $\Z{6}^{(\text{PG})}$                          & $k$ \\
$\left(0,\frac{1}{6},\frac{1}{3},-\frac{1}{2}\right)$
                     &         & $\Z{3} \times \left(\Z{2}\right)^2$            & $(n_3 + n_4, n_5, n_6)$ \\
\cline{2-4}
                     &2--1     & $\Z{6}^{(\text{PG})}$                          & $k$ \\
                     &         & $\Z{3} \times \left(\Z{2}\right)^2$            & $(n_1 + n_2 + n_3 + n_4, n_5, n_6)$ \\
\cline{2-4}
                     &3--1     & $\Z{6}^{(\text{PG})}$                          & $k$ \\
                     &         & $\Z{3} \times \left(\Z{2}\right)^2$            & $(n_4 + n_5, n_1 + n_2 + n_3, n_6)$ \\
\cline{2-4}
                     &4--1     & $\Z{6}^{(\text{PG})}$                          & $k$ \\
                     &         & $\Z{6} \times \Z{2}$                           & $(n_1 + n_2 + n_3 + n_4 + n_5, n_6)$ \\
\hline
$\Z{7}$              &1--1     & $\Z{7}^{(\text{PG})}$                          & $k$ \\
$\left(0,\frac{1}{7},\frac{2}{7},-\frac{3}{7}\right)$
                     &         & $\Z{7}$                                        & $n_1 + n_2 + n_3 + n_4 + n_5 + n_6$ \\
\hline
$\Z{8}$--I           &1--1     & $\Z{8}^{(\text{PG})}$                          & $k$ \\
$\left(0,\frac{1}{8},\frac{1}{4},-\frac{3}{8}\right)$
                     &         & $\left(\Z{2}\right)^2$                         & $(n_1 + n_2 + n_3 + n_4, n_5 + n_6)$ \\
\cline{2-4}
                     &2--1     & $\Z{8}^{(\text{PG})}$                          & $k$ \\
                     &         & $\left(\Z{2}\right)^2$                         & $(n_1 + n_2 + n_3 + n_4, n_5 + n_6)$ \\
\cline{2-4}
                     &3--1     & $\Z{8}^{(\text{PG})}$                          & $k$ \\
                     &         & $\Z{4}$                                        & $n_1 + n_2 + n_3 + n_4 + n_5 + n_6$ \\
\hline
$\Z{8}$--II          &1--1     & $\Z{8}^{(\text{PG})}$                          & $k$ \\
$\left(0,\frac{1}{8},\frac{3}{8},-\frac{1}{2}\right)$
                     &         & $\left(\Z{2}\right)^3$                         & $(n_1 + n_2 + n_3 + n_4, n_5, n_6)$ \\
\cline{2-4}
                     &2--1     & $\Z{8}^{(\text{PG})}$                          & $k$ \\
                     &         & $\Z{4} \times \Z{2}$                           & $(n_1 + n_2 + n_3 + n_4 + n_5, n_6)$ \\
\hline
$\Z{12}$--I          &1--1     & $\Z{12}^{(\text{PG})}$                         & $k$ \\
$\left(0,\frac{1}{12},\frac{1}{3},-\frac{5}{12}\right)$
                     &         & $\Z{3}$                                        & $n_5 + n_6$ \\
\cline{2-4}
                     &2--1     & $\Z{12}^{(\text{PG})}$                         & $k$ \\
                     &         & $\Z{3}$                                        & $n_1 + n_2 + n_3 + n_4 + n_5 + n_6$ \\
\hline
$\Z{12}$--II         &1--1     & $\Z{12}^{(\text{PG})}$                         & $k$ \\
$\left(0,\frac{1}{12},\frac{5}{12},-\frac{1}{2}\right)$
                     &         & $\left(\Z{2}\right)^2$                         & $(n_5, n_6)$ \\
\hline
$\Z{2}\times\Z{2}$   &1--1     & $\left(\Z{2}\times\Z{2}\right)^{(\text{PG})}$  & $(k,\ell)$ \\
$\left(0,0,\frac{1}{2},-\frac{1}{2}\right)$
                     &         & $\left(\Z{2}\right)^6$                         & $(n_1, n_2, n_3, n_4, n_5, n_6)$ \\
\cline{2-4}
$\left(0,\frac{1}{2},0,-\frac{1}{2}\right)$
                     &1--2     & $\left(\Z{2}\times\Z{2}\right)^{(\text{PG})}$  & $(k,\ell)$ \\
                     &         & $\left(\Z{2}\right)^5$                         & $(n_1, n_3, n_4, n_5, n_6)$ \\
\cline{2-4}
                     &1--3     & $\left(\Z{4}\times\Z{2}\right)^{(\text{PG})}$  & $(k+2(n_2+n_6), \ell)$ \\
                     &         & $\left(\Z{2}\right)^4$                         & $(n_1, n_3, n_4, n_5)$ \\
\cline{2-4}
                     &1--4     & $\left(\Z{4}\times\Z{4}\right)^{(\text{PG})}$  & $(k+2(n_2+n_6),\ell+2(n_4+n_6))$ \\
                     &         & $\left(\Z{2}\right)^3$                         & $(n_1, n_3, n_5)$ \\
\cline{2-4}
                     &2--1     & $\left(\Z{2}\times\Z{2}\right)^{(\text{PG})}$  & $(k,\ell)$ \\
                     &         & $\left(\Z{2}\right)^5$                         & $(n_1+n_2, n_3, n_4, n_5, n_6)$ \\
\cline{2-4}
                     &2--2     & $\left(\Z{2}\times\Z{2}\right)^{(\text{PG})}$  & $(k,\ell)$ \\
                     &         & $\left(\Z{2}\right)^4$                         & $(n_1+n_2, n_4, n_5, n_6)$ \\
\cline{2-4}
                     &2--3     & $\left(\Z{4}\times\Z{2}\right)^{(\text{PG})}$  & $(k+2(n_3+n_6), \ell)$ \\
                     &         & $\left(\Z{2}\right)^3$                         & $(n_1+n_2, n_4, n_5)$ \\
\cline{2-4}
                     &2--4     & $\left(\Z{2}\times\Z{2}\right)^{(\text{PG})}$  & $(k,\ell)$ \\
                     &         & $\left(\Z{2}\right)^4$                         & $(n_1+n_2, n_3, n_4, n_6)$ \\
\cline{2-4}
                     &2--5     & $\left(\Z{4}\times\Z{2}\right)^{(\text{PG})}$  & $(k+\ell+2(n_3+n_5),\ell)$ \\
                     &         & $\left(\Z{2}\right)^3$                         & $(n_1+n_2, n_4, n_6)$ \\
\cline{2-4}
                     &2--6     & $\left(\Z{4}\times\Z{4}\right)^{(\text{PG})}$  & $(k+2(n_3+n_6), \ell+2(n_5+n_6))$ \\
                     &         & $\left(\Z{2}\right)^2$                         & $(n_1+n_2, n_4)$ \\
\cline{2-4}
                     &3--1     & $\left(\Z{2}\times\Z{2}\right)^{(\text{PG})}$  & $(k,\ell)$ \\
                     &         & $\left(\Z{2}\right)^5$                         & $(n_1, n_2+n_3, n_4, n_5, n_6)$ \\
\cline{2-4}
                     &3--2     & $\left(\Z{2}\times\Z{2}\right)^{(\text{PG})}$  & $(k,\ell)$ \\
                     &         & $\left(\Z{2}\right)^4$                         & $(n_1, n_2+n_3, n_4, n_5)$ \\
\cline{2-4}
                     &3--3     & $\left(\Z{2}\times\Z{4}\right)^{(\text{PG})}$  & $(k, \ell+2(n_5+n_6))$ \\
                     &         & $\left(\Z{2}\right)^3$                         & $(n_1, n_2+n_3, n_4)$ \\
\cline{2-4}
                     &3--4     & $\left(\Z{4}\times\Z{4}\right)^{(\text{PG})}$  & $(k+2(n_4+n_6), \ell+2(n_5+n_6))$ \\
                     &         & $\left(\Z{2}\right)^2$                         & $(n_1, n_2+n_3)$ \\
\cline{2-4}
                     &4--1     & $\left(\Z{2}\times\Z{2}\right)^{(\text{PG})}$  & $(k,\ell)$ \\
                     &         & $\left(\Z{2}\right)^4$                         & $(n_1+n_2, n_3, n_4, n_5+n_6)$ \\
\cline{2-4}
                     &4--2     & $\left(\Z{2}\times\Z{2}\right)^{(\text{PG})}$  & $(k,\ell)$ \\
                     &         & $\left(\Z{2}\right)^3$                         & $(n_1+n_2, n_3, n_5+n_6)$ \\
\cline{2-4}
                     &5--1     & $\left(\Z{2}\times\Z{2}\right)^{(\text{PG})}$  & $(k,\ell)$ \\
                     &         & $\Z{4}\times\left(\Z{2}\right)^3$              & $(n_1+n_2+n_3, n_4, n_5, n_6)$ \\
\cline{2-4}
                     &5--2     & $\left(\Z{2}\times\Z{2}\right)^{(\text{PG})}$  & $(k,\ell)$ \\
                     &         & $\Z{4}\times\left(\Z{2}\right)^2$              & $(n_1+n_2+n_3, n_5, n_6)$ \\
\cline{2-4}
                     &5--3     & $\left(\Z{2}\times\Z{2}\right)^{(\text{PG})}$  & $(k,\ell)$ \\*
                     &         & $\left(\Z{2}\right)^4$                         & $(n_1+n_2+n_3, n_4, n_5, n_6)$ \\
\cline{2-4}
                     &5--4     & $\left(\Z{4}\times\Z{2}\right)^{(\text{PG})}$  & $(k+\ell+2(n_4+n_5),\ell)$ \\
                     &         & $\Z{4}\times\Z{2}$                             & $(n_1+n_2+n_3, n_6)$ \\
\cline{2-4}
                     &5--5     & $\left(\Z{4}\times\Z{4}\right)^{(\text{PG})}$  & $(k+2(n_4+n_6), \ell+2(n_5+n_6))$ \\
                     &         & $\Z{4}$                                        & $n_1+n_2+n_3$ \\
\cline{2-4}
                     &6--1     & $\left(\Z{2}\times\Z{2}\right)^{(\text{PG})}$  & $(k,\ell)$ \\
                     &         & $\left(\Z{2}\right)^4$                         & $(n_1+n_2, n_3+n_4, n_5, n_6)$ \\
\cline{2-4}
                     &6--2     & $\left(\Z{2}\times\Z{2}\right)^{(\text{PG})}$  & $(k,\ell)$ \\
                     &         & $\left(\Z{2}\right)^3$                         & $(n_1+n_2, n_3+n_4, n_6)$ \\
\cline{2-4}
                     &6--3     & $\left(\Z{2}\times\Z{4}\right)^{(\text{PG})}$  & $(k,\ell + 2(n_5+n_6))$ \\
                     &         & $\left(\Z{2}\right)^2$                         & $(n_1+n_2, n_3+n_4)$ \\
\cline{2-4}
                     &7--1     & $\left(\Z{2}\times\Z{2}\right)^{(\text{PG})}$  & $(k,\ell)$ \\
                     &         & $\left(\Z{2}\right)^4$                         & $(n_1+n_2, n_3, n_4+n_5, n_6)$ \\
\cline{2-4}
                     &7--2     & $\left(\Z{2}\times\Z{2}\right)^{(\text{PG})}$  & $(k,\ell)$ \\
                     &         & $\left(\Z{2}\right)^3$                         & $(n_1+n_2, n_3, n_4+n_5)$ \\
\cline{2-4}
                     &8--1     & $\left(\Z{2}\times\Z{2}\right)^{(\text{PG})}$  & $(k,\ell)$ \\
                     &         & $\left(\Z{2}\right)^4$                         & $(n_1, n_2+n_3, n_4, n_5+n_6)$ \\
\cline{2-4}
                     &9--1     & $\left(\Z{2}\times\Z{2}\right)^{(\text{PG})}$  & $(k,\ell)$ \\
                     &         & $\Z{4}\times\left(\Z{2}\right)^2$              & $(n_1+n_2+n_3, n_4+n_5, n_6)$ \\
\cline{2-4}
                     &9--2     & $\left(\Z{2}\times\Z{2}\right)^{(\text{PG})}$  & $(k,\ell)$ \\
                     &         & $\Z{4}\times\Z{2}$                             & $(n_1+n_2+n_3, n_4+n_5)$ \\
\cline{2-4}
                     &9--3     & $\left(\Z{2}\times\Z{2}\right)^{(\text{PG})}$  & $(k,\ell)$ \\
                     &         & $\left(\Z{2}\right)^3$                         & $(n_1+n_2+n_3, n_4+n_5, n_6)$ \\
\cline{2-4}
                     &10--1    & $\left(\Z{2}\times\Z{2}\right)^{(\text{PG})}$  & $(k,\ell)$ \\
                     &         & $\Z{4}\times\left(\Z{2}\right)^2$              & $(n_1+n_2+n_3, n_4, n_5+n_6)$ \\
\cline{2-4}
                     &10--2    & $\left(\Z{2}\times\Z{2}\right)^{(\text{PG})}$  & $(k,\ell)$ \\
                     &         & $\left(\Z{2}\right)^3$                         & $(n_1+n_2+n_3, n_4, n_5+n_6)$ \\
\cline{2-4}
                     &11--1    & $\left(\Z{2}\times\Z{2}\right)^{(\text{PG})}$  & $(k,\ell)$ \\
                     &         & $\left(\Z{2}\right)^3$                         & $(n_1+n_2, n_3+n_4, n_5+n_6)$ \\
\cline{2-4}
                     &12--1    & $\left(\Z{2}\times\Z{2}\right)^{(\text{PG})}$  & $(k,\ell)$ \\
                     &         & $\left(\Z{4}\right)^2$                         & $(n_1+n_2+n_3, n_4+n_5+n_6)$ \\
\cline{2-4}
                     &12--2    & $\left(\Z{2}\times\Z{2}\right)^{(\text{PG})}$  & $(k,\ell)$ \\
                     &         & $\Z{4}\times\Z{2}$                             & $(n_1+n_2+n_3, n_4+n_5+n_6)$ \\
\hline
$\Z{2}\times\Z{4}$   &1--1     & $\left(\Z{2}\times\Z{4}\right)^{(\text{PG})}$  & $(k,\ell)$ \\
$\left(0,0,\frac{1}{2},-\frac{1}{2}\right)$
                     &         & $\left(\Z{2}\right)^4$                         & $(n_1+n_2, n_3, n_4, n_5+n_6)$ \\
\cline{2-4}
$\left(0,\frac{1}{4},0,-\frac{1}{4}\right)$
                     &1--2     & $\left(\Z{2}\times\Z{4}\right)^{(\text{PG})}$  & $(k,\ell)$ \\
                     &         & $\left(\Z{2}\right)^3$                         & $(n_3, n_4, n_5+n_6)$ \\
\cline{2-4}
                     &1--3     & $\left(\Z{2}\times\Z{4}\right)^{(\text{PG})}$  & $(k,\ell)$ \\*
                     &         & $\left(\Z{2}\right)^3$                         & $(n_1+n_2+n_5+n_6, n_3, n_4)$ \\
\cline{2-4}
                     &1--4     & $\left(\Z{2}\times\Z{4}\right)^{(\text{PG})}$  & $(k,\ell)$ \\
                     &         & $\left(\Z{2}\right)^3$                         & $(n_1+n_2, n_3, n_5+n_6)$ \\
\cline{2-4}
                     &1--5     & $\left(\Z{2}\times\Z{4}\right)^{(\text{PG})}$  & $(k,\ell)$ \\
                     &         & $\left(\Z{2}\right)^3$                         & $(n_1+n_2+n_4, n_3, n_5+n_6)$ \\
\cline{2-4}
                     &1--6     & $\left(\Z{2}\times\Z{4}\right)^{(\text{PG})}$  & $(k,\ell)$ \\
                     &         & $\left(\Z{2}\right)^3$                         & $(n_1+n_2+n_4, n_3, n_4+n_5+n_6)$ \\
\cline{2-4}
                     &2--1     & $\left(\Z{2}\times\Z{4}\right)^{(\text{PG})}$  & $(k,\ell)$ \\
                     &         & $\left(\Z{2}\right)^4$                         & $(n_1+n_2, n_3+n_4, n_5, n_6)$ \\
\cline{2-4}
                     &2--2     & $\left(\Z{2}\times\Z{4}\right)^{(\text{PG})}$  & $(k,\ell)$ \\
                     &         & $\left(\Z{2}\right)^3$                         & $(n_1+n_2, n_3+n_4, n_5)$ \\
\cline{2-4}
                     &2--3     & $\left(\Z{4}\times\Z{4}\right)^{(\text{PG})}$  & $(k+2(n_1+n_2+n_3+n_4),\ell)$ \\
                     &         & $\left(\Z{2}\right)^2$                         & $(n_5, n_6)$ \\
\cline{2-4}
                     &2--4     & $\left(\Z{4}\times\Z{4}\right)^{(\text{PG})}$  & $(k+2(n_3+n_4+n_6),\ell)$ \\
                     &         & $\left(\Z{2}\right)^2$                         & $(n_1+n_2+n_6, n_5)$ \\
\cline{2-4}
                     &2--5     & $\left(\Z{2}\times\Z{4}\right)^{(\text{PG})}$  & $(k,\ell)$ \\
                     &         & $\left(\Z{2}\right)^3$                         & $(n_1+n_2, n_5, n_6)$ \\
\cline{2-4}
                     &2--6     & $\left(\Z{2}\times\Z{4}\right)^{(\text{PG})}$  & $(k,\ell)$ \\
                     &         & $\left(\Z{2}\right)^3$                         & $(n_1+n_2, n_3+n_4+n_6, n_5)$ \\
\cline{2-4}
                     &3--1     & $\left(\Z{2}\times\Z{4}\right)^{(\text{PG})}$  & $(k,\ell)$ \\
                     &         & $\left(\Z{2}\right)^3$                         & $(n_1+n_2+n_3, n_4, n_5+n_6)$ \\
\cline{2-4}
                     &3--2     & $\left(\Z{2}\times\Z{4}\right)^{(\text{PG})}$  & $(k,\ell)$ \\
                     &         & $\left(\Z{2}\right)^2$                         & $(n_1+n_2+n_3, n_4)$ \\
\cline{2-4}
                     &3--3     & $\left(\Z{2}\times\Z{4}\right)^{(\text{PG})}$  & $(k,\ell)$ \\
                     &         & $\left(\Z{2}\right)^2$                         & $(n_1+n_2+n_3, n_5+n_6)$ \\
\cline{2-4}
                     &3--4     & $\left(\Z{2}\times\Z{4}\right)^{(\text{PG})}$  & $(k,\ell)$ \\
                     &         & $\left(\Z{2}\right)^2$                         & $(n_1+n_2+n_3, n_4+n_5+n_6)$ \\
\cline{2-4}
                     &3--5     & $\left(\Z{2}\times\Z{4}\right)^{(\text{PG})}$  & $(k,\ell)$ \\
                     &         & $\left(\Z{2}\right)^2$                         & $(n_4, n_5+n_6)$ \\
\cline{2-4}
                     &3--6     & $\left(\Z{2}\times\Z{4}\right)^{(\text{PG})}$  & $(k,\ell)$ \\
                     &         & $\left(\Z{2}\right)^2$                         & $(n_1+n_2+n_3+n_5+n_6, n_4)$ \\
\cline{2-4}
                     &4--1     & $\left(\Z{2}\times\Z{4}\right)^{(\text{PG})}$  & $(k,\ell)$ \\
                     &         & $\left(\Z{2}\right)^3$                         & $(n_1+n_2, n_3+n_4+n_5, n_6)$ \\
\cline{2-4}
                     &4--2     & $\left(\Z{2}\times\Z{4}\right)^{(\text{PG})}$  & $(k,\ell)$ \\
                     &         & $\left(\Z{2}\right)^2$                         & $(n_1+n_2, n_3+n_4+n_5)$ \\
\cline{2-4}
                     &4--3     & $\left(\Z{2}\times\Z{4}\right)^{(\text{PG})}$  & $(k,\ell)$ \\
                     &         & $\left(\Z{2}\right)^2$                         & $(n_1+n_2+n_3+n_4+n_5, n_6)$ \\
\cline{2-4}
                     &4--4     & $\left(\Z{2}\times\Z{4}\right)^{(\text{PG})}$  & $(k,\ell)$ \\*
                     &         & $\left(\Z{2}\right)^2$                         & $(n_1+n_2+n_6, n_3+n_4+n_5+n_6)$ \\
\cline{2-4}
                     &4--5     & $\left(\Z{2}\times\Z{4}\right)^{(\text{PG})}$  & $(k,\ell)$ \\
                     &         & $\left(\Z{2}\right)^2$                         & $(n_1+n_2, n_6)$ \\
\cline{2-4}
                     &5--1     & $\left(\Z{2}\times\Z{4}\right)^{(\text{PG})}$  & $(k,\ell)$ \\
                     &         & $\left(\Z{2}\right)^3$                         & $(n_1+n_2+n_3+n_4, n_5, n_6)$ \\
\cline{2-4}
                     &5--2     & $\left(\Z{2}\times\Z{4}\right)^{(\text{PG})}$  & $(k,\ell)$ \\
                     &         & $\left(\Z{2}\right)^2$                         & $(n_1+n_2+n_3+n_4, n_5)$ \\
\cline{2-4}
                     &6--1     & $\left(\Z{2}\times\Z{4}\right)^{(\text{PG})}$  & $(k,\ell)$ \\
                     &         & $\left(\Z{2}\right)^3$                         & $(n_1+n_2+n_4+n_5, n_3, n_6)$ \\
\cline{2-4}
                     &6--2     & $\left(\Z{2}\times\Z{4}\right)^{(\text{PG})}$  & $(k,\ell)$ \\
                     &         & $\left(\Z{2}\right)^2$                         & $(n_3, n_6)$ \\
\cline{2-4}
                     &6--3     & $\left(\Z{2}\times\Z{4}\right)^{(\text{PG})}$  & $(k,\ell)$ \\
                     &         & $\left(\Z{2}\right)^2$                         & $(n_1+n_2+n_4+n_5, n_3)$ \\
\cline{2-4}
                     &6--4     & $\left(\Z{2}\times\Z{4}\right)^{(\text{PG})}$  & $(k,\ell)$ \\
                     &         & $\left(\Z{2}\right)^2$                         & $(n_1+n_2+n_4+n_5+n_6, n_3)$ \\
\cline{2-4}
                     &6--5     & $\left(\Z{4}\times\Z{4}\right)^{(\text{PG})}$  & $(k+2(n_1+n_2+n_3+n_4+n_5),\ell)$ \\
                     &         & $\Z{2}$                                        & $n_6$ \\
\cline{2-4}
                     &7--1     & $\left(\Z{2}\times\Z{4}\right)^{(\text{PG})}$  & $(k,\ell)$ \\
                     &         & $\left(\Z{2}\right)^3$                         & $(n_1+n_2+n_3+n_4, n_5, n_6)$ \\
\cline{2-4}
                     &7--2     & $\left(\Z{2}\times\Z{4}\right)^{(\text{PG})}$  & $(k,\ell)$ \\
                     &         & $\left(\Z{2}\right)^2$                         & $(n_1+n_2+n_3+n_4, n_5)$ \\
\cline{2-4}
                     &7--3     & $\left(\Z{2}\times\Z{4}\right)^{(\text{PG})}$  & $(k,\ell)$ \\
                     &         & $\left(\Z{2}\right)^2$                         & $(n_1+n_2+n_3+n_4, n_6)$ \\
\cline{2-4}
                     &8--1     & $\left(\Z{2}\times\Z{4}\right)^{(\text{PG})}$  & $(k,\ell)$ \\
                     &         & $\left(\Z{2}\right)^2$                         & $(n_1+n_2+n_3, n_4+n_5+n_6)$ \\
\cline{2-4}
                     &8--2     & $\left(\Z{2}\times\Z{4}\right)^{(\text{PG})}$  & $(k,\ell)$ \\
                     &         & $\Z{2}$                                        & $n_4+n_5+n_6$ \\
\cline{2-4}
                     &8--3     & $\left(\Z{2}\times\Z{4}\right)^{(\text{PG})}$  & $(k,\ell)$ \\
                     &         & $\Z{2}$                                        & $n_1+n_2+n_3+n_4+n_5+n_6$ \\
\cline{2-4}
                     &9--1     & $\left(\Z{2}\times\Z{4}\right)^{(\text{PG})}$  & $(k,\ell)$ \\
                     &         & $\Z{4}\times\Z{2}$                             & $(n_1+n_2+n_3+n_4+n_5, n_6)$ \\
\cline{2-4}
                     &9--2     & $\left(\Z{2}\times\Z{4}\right)^{(\text{PG})}$  & $(k,\ell)$ \\
                     &         & $\Z{4}$                                        & $n_1+n_2+n_3+n_4+n_5$ \\
\cline{2-4}
                     &9--3     & $\left(\Z{2}\times\Z{4}\right)^{(\text{PG})}$  & $(k,\ell)$ \\
                     &         & $\left(\Z{2}\right)^2$                         & $(n_1+n_2+n_3+n_4+n_5, n_6)$ \\
\cline{2-4}
                     &10--1    & $\left(\Z{2}\times\Z{4}\right)^{(\text{PG})}$  & $(k,\ell)$ \\
                     &         & $\left(\Z{2}\right)^2$                         & $(n_1+n_2+n_3+n_4+n_5, n_6)$ \\
\cline{2-4}
                     &10--2    & $\left(\Z{2}\times\Z{4}\right)^{(\text{PG})}$  & $(k,\ell)$ \\*
                     &         & $\Z{2}$                                        & $n_1+n_2+n_3+n_4+n_5$ \\
\hline
$\Z{2}\times\Z{6}$--I &1--1    & $\left(\Z{2}\times\Z{6}\right)^{(\text{PG})}$  & $(k,\ell)$ \\
$\left(0,0,\frac{1}{2},-\frac{1}{2}\right)$
                     &         & $\left(\Z{2}\right)^2$                         & $(n_3, n_4)$ \\
\cline{2-4}
$\left(0,\frac{1}{6},0,-\frac{1}{6}\right)$
                     &1--2     & $\left(\Z{2}\times\Z{6}\right)^{(\text{PG})}$  & $(k,\ell)$ \\
                     &         & $\Z{2}$                                        & $n_3$ \\
\cline{2-4}
                     &2--1     & $\left(\Z{2}\times\Z{6}\right)^{(\text{PG})}$  & $(k,\ell)$ \\
                     &         & $\left(\Z{2}\right)^2$                         & $(n_5, n_6)$ \\
\cline{2-4}
                     &2--2     & $\left(\Z{2}\times\Z{6}\right)^{(\text{PG})}$  & $(k,\ell)$ \\
                     &         & $\Z{2}$                                        & $n_5$ \\
\hline
$\Z{2}\times\Z{6}$--II &1--1   & $\left(\Z{2}\times\Z{6}\right)^{(\text{PG})}$  & $(k,\ell)$ \\
\cline{2-4}
$\left(0,0,\frac{1}{2},-\frac{1}{2}\right)$
                     &2--1     & $\left(\Z{2}\times\Z{6}\right)^{(\text{PG})}$  & $(k,\ell)$ \\
\cline{2-4}
$\left(0,\frac{1}{6},\frac{1}{6},-\frac{1}{3}\right)$
                     &3--1     & $\left(\Z{2}\times\Z{6}\right)^{(\text{PG})}$  & $(k,\ell)$ \\
\cline{2-4}
                     &4--1     & $\left(\Z{2}\times\Z{6}\right)^{(\text{PG})}$  & $(k,\ell)$ \\
\hline
$\Z{3}\times\Z{3}$   &1--1     & $\left(\Z{3}\times\Z{3}\right)^{(\text{PG})}$  & $(k,\ell)$ \\
$\left(0,0,\frac{1}{3},-\frac{1}{3}\right)$
                     &         & $\left(\Z{3}\right)^3$                         & $(n_1+n_2, n_3+n_4, n_5+n_6)$ \\
\cline{2-4}
$\left(0,\frac{1}{3},0,-\frac{1}{3}\right)$
                     &1--2     & $\left(\Z{3}\times\Z{3}\right)^{(\text{PG})}$  & $(k,\ell)$ \\
                     &         & $\left(\Z{3}\right)^2$                         & $(n_1+n_2, n_3+n_4)$ \\
\cline{2-4}
                     &1--3     & $\left(\Z{3}\times\Z{3}\right)^{(\text{PG})}$  & $(k,\ell)$ \\
                     &         & $\left(\Z{3}\right)^2$                         & $(2(n_1+n_2)+n_5+n_6, n_3+n_4)$ \\
\cline{2-4}
                     &1--4     & $\left(\Z{3}\times\Z{3}\right)^{(\text{PG})}$  & $(k,\ell)$ \\
                     &         & $\left(\Z{3}\right)^2$                         & $(n_1+n_2+2(n_3+n_4), 2(n_3+n_4)+n_5+n_6)$ \\
\cline{2-4}
                     &2--1     & $\left(\Z{3}\times\Z{3}\right)^{(\text{PG})}$  & $(k,\ell)$ \\
                     &         & $\left(\Z{3}\right)^2$                         & $(n_1+n_2+n_3+n_4, n_5+n_6)$ \\
\cline{2-4}
                     &2--2     & $\left(\Z{3}\times\Z{3}\right)^{(\text{PG})}$  & $(k,\ell)$ \\
                     &         & $\Z{3}$                                        & $n_1+n_2+n_3+n_4$ \\
\cline{2-4}
                     &2--3     & $\left(\Z{3}\times\Z{3}\right)^{(\text{PG})}$  & $(k,\ell)$ \\
                     &         & $\Z{3}$                                        & $n_5+n_6$ \\
\cline{2-4}
                     &2--4     & $\left(\Z{3}\times\Z{3}\right)^{(\text{PG})}$  & $(k,\ell)$ \\
                     &         & $\Z{3}$                                        & $n_1+n_2+n_3+n_4+2(n_5+n_6)$ \\
\cline{2-4}
                     &3--1     & $\left(\Z{3}\times\Z{3}\right)^{(\text{PG})}$  & $(k,\ell)$ \\
                     &         & $\left(\Z{3}\right)^2$                         & $(n_1+n_2+n_3+n_4, n_5+n_6)$ \\
\cline{2-4}
                     &3--2     & $\left(\Z{3}\times\Z{3}\right)^{(\text{PG})}$  & $(k,\ell)$ \\
                     &         & $\Z{3}$                                        & $n_5+n_6$ \\
\cline{2-4}
                     &3--3     & $\left(\Z{3}\times\Z{3}\right)^{(\text{PG})}$  & $(k,\ell)$ \\
                     &         & $\Z{3}$                                        & $n_1+n_2+n_3+n_4+n_5+n_6$ \\
\cline{2-4}
                     &4--1     & $\left(\Z{3}\times\Z{3}\right)^{(\text{PG})}$  & $(k,\ell)$ \\
                     &         & $\left(\Z{3}\right)^2$                         & $(n_1+n_4, n_2+n_3+n_5+n_6)$ \\
\cline{2-4}
                     &4--2     & $\left(\Z{3}\times\Z{3}\right)^{(\text{PG})}$  & $(k,\ell)$ \\*
                     &         & $\Z{3}$                                        & $n_1+n_4$ \\
\cline{2-4}
                     &4--3     & $\left(\Z{9}\times\Z{3}\right)^{(\text{PG})}$  & $(k+2\ell+6(n_1+n_2+n_3+n_4+n_5+n_6),\ell)$ \\
\cline{2-4}
                     &5--1     & $\left(\Z{3}\times\Z{3}\right)^{(\text{PG})}$  & $(k,\ell)$ \\
                     &         & $\Z{3}$                                        & $n_1+n_2+n_3+n_4+n_5+n_6$ \\
\hline
$\Z{3}\times\Z{6}$   &1--1     & $\left(\Z{3}\times\Z{6}\right)^{(\text{PG})}$  & $(k,\ell)$ \\
$\left(0,0,\frac{1}{3},-\frac{1}{3}\right)$
                     &         & $\Z{3}$                                        & $n_3+n_4$ \\
\cline{2-4}
$\left(0,\frac{1}{6},0,-\frac{1}{6}\right)$
                     &1--2     & $\left(\Z{3}\times\Z{6}\right)^{(\text{PG})}$  & $(k,\ell)$ \\
\cline{2-4}
                     &2--1     & $\left(\Z{3}\times\Z{6}\right)^{(\text{PG})}$  & $(k,\ell)$ \\
                     &         & $\Z{3}$                                        & $n_5+n_6$ \\
\cline{2-4}
                     &2--2     & $\left(\Z{3}\times\Z{6}\right)^{(\text{PG})}$  & $(k,\ell)$ \\
\hline
$\Z{4}\times\Z{4}$   &1--1     & $\left(\Z{4}\times\Z{4}\right)^{(\text{PG})}$  & $(k,\ell)$ \\
$\left(0,0,\frac{1}{4},-\frac{1}{4}\right)$ 
                     &         & $\left(\Z{2}\right)^3$                         & $(n_1+n_2, n_3+n_4,n_5+n_6)$ \\
\cline{2-4}
$\left(0,\frac{1}{4},0,-\frac{1}{4}\right)$
                     &1--2     & $\left(\Z{4}\times\Z{4}\right)^{(\text{PG})}$  & $(k,\ell)$ \\
                     &         & $\left(\Z{2}\right)^2$                         & $(n_1+n_2, n_3+n_4)$ \\
\cline{2-4}
                     &1--3     & $\left(\Z{4}\times\Z{4}\right)^{(\text{PG})}$  & $(k,\ell)$ \\
                     &         & $\left(\Z{2}\right)^2$                         & $(n_1+n_2+n_5+n_6, n_3+n_4)$ \\
\cline{2-4}
                     &1--4     & $\left(\Z{4}\times\Z{4}\right)^{(\text{PG})}$  & $(k,\ell)$ \\
                     &         & $\left(\Z{2}\right)^2$                         & $(n_1+n_2+n_3+n_4, n_3+n_4+n_5+n_6)$ \\
\cline{2-4}
                     &2--1     & $\left(\Z{4}\times\Z{4}\right)^{(\text{PG})}$  & $(k,\ell)$ \\
                     &         & $\left(\Z{2}\right)^2$                         & $(n_1+n_2+n_3+n_4, n_5+n_6)$ \\
\cline{2-4}
                     &2--2     & $\left(\Z{4}\times\Z{4}\right)^{(\text{PG})}$  & $(k,\ell)$ \\
                     &         & $\Z{2}$                                        & $n_5+n_6$ \\
\cline{2-4}
                     &2--3     & $\left(\Z{4}\times\Z{4}\right)^{(\text{PG})}$  & $(k,\ell)$ \\
                     &         & $\Z{2}$                                        & $n_1+n_2+n_3+n_4$ \\
\cline{2-4}
                     &2--4     & $\left(\Z{4}\times\Z{4}\right)^{(\text{PG})}$  & $(k,\ell)$ \\
                     &         & $\Z{2}$                                        & $n_1+n_2+n_3+n_4+n_5+n_6$ \\
\cline{2-4}
                     &3--1     & $\left(\Z{4}\times\Z{4}\right)^{(\text{PG})}$  & $(k,\ell)$ \\
                     &         & $\left(\Z{2}\right)^2$                         & $(n_1+n_2, n_3+n_4+n_5+n_6)$ \\
\cline{2-4}
                     &3--2     & $\left(\Z{4}\times\Z{4}\right)^{(\text{PG})}$  & $(k,\ell)$ \\
                     &         & $\Z{2}$                                        & $n_1+n_2+n_3+n_4+n_5+n_6$ \\
\cline{2-4}
                     &4--1     & $\left(\Z{4}\times\Z{4}\right)^{(\text{PG})}$  & $(k,\ell)$ \\
                     &         & $\left(\Z{2}\right)^2$                         & $(n_1+n_2, n_3+n_4+n_5+n_6)$ \\
\cline{2-4}
                     &4--2     & $\left(\Z{4}\times\Z{4}\right)^{(\text{PG})}$  & $(k,\ell)$ \\
                     &         & $\Z{2}$                                        & $n_3+n_4+n_5+n_6$ \\
\cline{2-4}
                     &4--3     & $\left(\Z{4}\times\Z{4}\right)^{(\text{PG})}$  & $(k,\ell)$ \\
                     &         & $\Z{2}$                                        & $n_1+n_2$ \\
\cline{2-4}
                     &5--1     & $\left(\Z{4}\times\Z{4}\right)^{(\text{PG})}$  & $(k,\ell)$ \\
                     &         & $\Z{2}$                                        & $n_1+n_2+n_3+n_4+n_5+n_6$ \\
\cline{2-4}
                     &5--2     & $\left(\Z{4}\times\Z{4}\right)^{(\text{PG})}$  & $(k,\ell)$ \\
\hline
$\Z{6}\times\Z{6}$   &1--1     & $\left(\Z{6}\times\Z{6}\right)^{(\text{PG})}$  & $(k,\ell)$ \\
$\left(0,0,\frac{1}{6},-\frac{1}{6}\right)$ 
                     &         &                                                & \\
$\left(0,\frac{1}{6},0,-\frac{1}{6}\right)$
                     &         &                                                & \\
\hline
\end{longtable}

\subsection[Discrete anomalies of the SG flavor symmetry $D_S$]{\boldmath Discrete anomalies of the SG flavor symmetry $D_S$\unboldmath}

Discrete groups can be anomalous and their anomaly coefficients can be computed directly using 
Fujikawa's method for the Jacobian of the path integral measure~\cite{Fujikawa:1979ay,Fujikawa:1980eg}. 
Following ref.~\cite{Araki:2007zza,Araki:2008ek}, the mixed $\Z{N}-G_i-G_i$ anomaly coefficient of 
a discrete group $\Z{N}$ and a non-Abelian gauge group factor $G_i$ is given by 
\begin{equation}
A_{\Z{N}-G_i-G_i} ~=~ \sum_{\rep{r}^{(f)}}\,q^{(f)}\,\ell\left(\rep{r}^{(f)}\right)\;.
\end{equation}
Here, the summation runs over all fermions that transform in the representation $\rep{r}^{(f)}$ of 
the gauge group factor $G_i$ and $q^{(f)}$ denotes their discrete $\Z{N}$ charge as given in 
table~\ref{tab:results} for a string with constructing element\footnote{Note that in the presence 
of roto-translations this constructing element $g$ is equal to 
$g = (\theta^k\,\omega^\ell, \lambda_{(k,\ell)} + n_i e_i)$ for some non-trivial translation 
$\lambda_{(k,\ell)}$. The discrete charges $q^{(f)}$, however, do not depend on $\lambda_{(k,\ell)}$ but only on 
the integers $n_i, k, \ell \in \Z{}$.}
\begin{equation}
g ~=~ \left(T_1\right)^{n_1}\,\ldots \left(T_6\right)^{n_6}\, \left(g_\theta\right)^k\, \left(g_\omega\right)^\ell ~\in~ S\;,
\end{equation}
with $n_1, \ldots, n_6, k, \ell \in \Z{}$. In addition, the 
Dynkin index $\ell\left(\rep{r}^{(f)}\right)$ is normalized such that $\ell(\rep{N}) = 1$ for the fundamental 
representation $\rep{N}$ of $\SU{N}$ and $\ell(\rep{2N}) = 2$ for the vector representation 
$\rep{2N}$ of $\SO{2N}$. Furthermore, the $\Z{N}-\text{grav.}-\text{grav.}$ anomaly coefficient reads
\begin{equation}
A_{\Z{N}-\text{grav.}-\text{grav.}} ~=~ \sum_m q^{(m)}\, \text{dim}(\rep{R}^{(m)})\;,
\end{equation}
where the summation runs over all fermions that transform in the representation $\rep{R}^{(m)}$ of 
$G_1 \times G_2 \times \ldots$, being the full non-Abelian gauge group of the theory.

For all space groups from table~\ref{tab:results} we have constructed more than 1,000 random 
orbifold models using the \texttt{orbifolder}~\cite{Nilles:2011aj}. For every orbifold model we have checked 
that for each $\Z{N}$ factor of $D_S$ there is a discrete Green-Schwarz constant 
$\Delta_\text{GS}$~\cite{Green:1984sg,Araki:2007zza} such that for all non-Abelian gauge group factors $G_i$ the 
anomalies are universal, i.e.
\begin{subequations}
\begin{eqnarray}
A_{\Z{N}-G_i-G_i}                   & = & \Delta_\text{GS} \;\text{ mod }\; N\;,\\
A_{\Z{N}-\text{grav.}-\text{grav.}} & = & 12\,A_{\Z{N}-G_i-G_i}  \;\text{ mod }\; N\;.
\end{eqnarray}
\end{subequations}
Consequently, if $\Delta_\text{GS} \neq 0$ the universal discrete anomalies can be canceled by a 
discrete Green-Schwarz mechanism involving a single, universally coupled axion. Thus, we have 
performed a non-trivial test of all SG flavor symmetries.

\section{Conclusions}

There are large sets of semi-realistic string orbifold models, whose phenomenology may reveal
interesting features of the string landscape, restricting thereby string constructions
in the search of an ultraviolet completion of low-energy physics. Studying the 
phenomenology of orbifold compactifications requires, as one of the first steps, the identification
and understanding of all symmetries of a string model. 

In this work, we have studied the Abelian SG flavor symmetry $D_S$ that arises from the 
constraints on closed strings to split and join while propagating on a six-dimensional orbifold, 
defined by a space group $S$. By demanding that $D_S$ be a (non-faithful) representation of $S$ and 
that the discrete charges of all closed strings be well-defined and conserved, we find that 
the SG flavor symmetry $D_S$ corresponds to the Abelianization of the space group $S$. We have 
discussed how this observation can be applied to space groups with different features, rendering 
the precise structure of $D_S$ and their charges for all closed strings.

In section~\ref{sec:results}, we computed $D_S$ for all 138 space groups with Abelian point group 
that yield $\mathcal{N}=1$ effective field theories in four dimensions. It is known that the 
identified symmetries, displayed in table~\ref{tab:results}, play a key role as part of the flavor 
symmetries at low energies and are thus essential for phenomenology. As a cross check of the 
validity of these symmetries, we have also explicitly verified in thousands of $\E8\times\E8$ 
heterotic orbifold models that all identified $D_S$ lead to universal anomalies, which allows them 
to be canceled by a universal Green-Schwarz mechanism.

The SG flavor symmetries $D_S$ are respected in interactions of both massless and massive strings. 
Interestingly, there are cases where the charges of massless strings are restricted such that the 
symmetry of the massless sector is only a subgroup of $D_S$. An intriguing consequence is that the 
lightest massive string (which can be a winding mode) can only be produced and 
annihilated in pairs. Thus, the lightest massive string is stable and, hence, contributes to a 
dark sector of the effective model, see section~\ref{sec:Z2xZ2Example51}. Whether this feature 
might be seen as an explanation of (some of) the dark matter of the universe and its evolution, or 
whether it rules out some of the orbifold geometries shall be studied elsewhere.

A natural extension of our work is the study of non-Abelian orbifolds{, i.e.\ orbifolds whose point 
groups $P$ are non-Abelian~\cite{Kakushadze:1996hj,Konopka:2012gy,Fischer:2013qza}. One should 
explore how the constraints we imposed on the SG flavor symmetries $D_S$ and their charges} apply 
to orbifolds with non-Abelian point groups, to answer whether this contribution to flavor 
symmetries is also Abelian in those scenarios.

\appendix

\section{Space groups and orbifolds}
\label{app:SpaceGroupsAndOrbifolds}

In this appendix we give a detailed review on space groups and their resulting orbifold geometries. 
We highlight an unconventional approach to use a so-called presentation of a space group, i.e.\ 
a way to define a space group by abstract generators and a set of relations among them without 
explicitly writing out neither a basis of lattice vectors nor the rotation matrices. In addition, 
we briefly discuss closed strings on orbifolds with a focus on those properties that are relevant 
for the space group selection rule of interacting strings.

\subsection{The space group}
\label{app:SpaceGroup}

A general element $g$ of a $D$-dimensional space group $S$ can be written as
\begin{equation}
g ~=~ (\rho, \lambda) ~\in~ S\;.
\end{equation}
We will mostly consider the case $D=6$ in order to compactify the heterotic string from ten to four 
dimensions. By definition, the space group element $g$ acts on the $D$ extra-dimensional 
coordinates $y \in\mathbbm{R}^D$ as
\begin{equation}
y ~\stackrel{g}{\mapsto}~ (\rho, \lambda)\, y ~=~ \rho\, y + \lambda\;,
\end{equation}
where the so-called twist $\rho \in \mathrm{O}(D)$ is a $D \times D$ rotation matrix (if 
det$(\rho)=1$) or reflection matrix (if det$(\rho)=-1$) and the vector $\lambda \in \mathbbm{R}^D$ 
yields a translation. Consequently, two space group elements $g_1 = (\rho_1, \lambda_1)$ and 
$g_2 = (\rho_2, \lambda_2)$ multiply as
\begin{equation}
(\rho_1, \lambda_1)\,(\rho_2, \lambda_2) ~=~ (\rho_1\,\rho_2, \rho_1\,\lambda_2 + \lambda_1)\;.
\end{equation}
It follows that $(\Id, 0) = \Id_S$ is the identity element of $S$ and the inverse element of 
$(\rho, \lambda)$ is given by
\begin{equation}
(\rho, \lambda)^{-1} ~=~ (\rho^{-1}, -\rho^{-1}\,\lambda)\;.
\end{equation}

Now we can define a $D$-dimensional space group $S$ by specifying finitely many generators: first, 
one chooses a $D$-dimensional torus lattice $\Lambda_D$, which is generated by $D$ linear independent 
translations $T_i$, i.e.\
\begin{equation}
T_i ~=~ (\Id, e_i) \qquad\text{for}\qquad i = 1,\ldots, D\;,
\end{equation}
where the basis vectors $e_i$ are given by the columns of a vielbein $e$. In addition, there are 
generators of the form 
\begin{equation}\label{eq:defrototranslation}
(\rho, \lambda) ~\in~ S \quad\text{with}\quad \rho \neq\Id \;.
\end{equation}
As a remark, if $\lambda \not\in \Lambda_D$ in eq.~\eqref{eq:defrototranslation} such generators are 
referred to as roto-translations. On the other hand, if $\lambda \in \Lambda_D$ one can choose a 
pure rotation as an alternative generator to eq.~\eqref{eq:defrototranslation}, i.e.
\begin{equation}
(\rho, 0) ~\in~ S\;.
\end{equation}
As we will be dealing with space groups with at most two rotational generators, we will label them 
by $g_\theta, g_\omega \in S$, where
\begin{equation}
g_\theta ~=~ (\theta, \lambda_\theta) \quad\text{and}\quad g_\omega ~=~ (\omega, \lambda_\omega)\;.
\end{equation}

The space group $S$ must close under multiplication. For example,
\begin{equation}\label{eqn:conjugatetranslation}
(\rho, \lambda)\, (\Id, e_i)\,(\rho, \lambda)^{-1} ~=~ (\Id, \rho\, e_i) ~\stackrel{!}{\in}~ S
\end{equation}
and, consequently, $\rho\, e_i$ must be from the $D$-dimensional lattice $\Lambda_D$. In other words, 
the twist $\rho$ of any space group element has to be an automorphism of the lattice 
$\Lambda_D$. Thus, one can always find a matrix $\hat\rho$ from $\text{GL}(D,\Z{})$ such that
\begin{equation}\label{eqn:TwistInLatticeBasis}
\rho\, e ~=~ e\, \hat\rho\;.
\end{equation}
$\hat\rho = e^{-1}\,\rho\, e$ is called the twist in the lattice basis. Since 
$\rho\in\mathrm{O}(D)$, we find the condition
\begin{equation}
\hat\rho^T G\, \hat\rho ~=~ G\;,
\end{equation}
on the $D$-dimensional torus metric $G=e^T\,e$.

The twists $\rho$ form a finite group $P$, called the point group. Its elements act 
crystallographically: they map the lattice $\Lambda_D$ to itself. As $P$ is finite, each element 
$\rho \in P$ has to have finite order, i.e.\ there exists a smallest integer 
$N_\rho \in \mathbbm{N}$ such that
\begin{equation}
\rho^{N_\rho} ~=~ \Id\;.
\end{equation}
A point group can be both, Abelian or non-Abelian. If $P$ is Abelian, it is isomorphic to the cyclic 
group (or to the direct product of several cyclic groups). In the following we restrict ourselves to 
Abelian point groups which preserve $\mathcal{N}=1$ supersymmetry, being either $\Z{M}$ or 
$\Z{M}\times\Z{N}$.

\subsection{Presentation of the space group}
\label{app:PresentationOfS}

The commutator of two space group elements $g,h \in S$ is defined as
\begin{equation}
[g,h] ~=~ g^{-1}h^{-1}g\,h\;.
\end{equation}
Then, since we restrict ourselves to Abelian point groups, the commutator of any pair of elements 
$g,h \in S$ is always a pure translation $\lambda_{(g, h)} \in \Lambda_D$, i.e.
\begin{equation}
[g,h] ~=~ (\Id, \lambda_{(g, h)})\;.
\end{equation}

Following ref.~\cite{Ratcliffe:2009}, we can specify a space group $S$ uniquely by a presentation 
that involves all relations between all generators. For example, in the case of a $\Z{M}\times\Z{N}$ 
point group we have the following generators of $S$: the rotations (or roto-translations) 
$g_\theta = (\theta, \lambda_\theta)$ and $g_\omega = (\omega, \lambda_\omega)$, 
and the translations $T_i = (\Id, e_i)$, $i=1,\ldots,D$. Then, all relations are given by
\begin{subequations}\label{eqn:SpaceGroupRelations}
\begin{eqnarray}
g_\theta^{M}            & = & \left(T_1\right)^{a_{(\theta,1)}}\, \ldots\, \left(T_D\right)^{a_{(\theta,D)}}\;,\label{eqn:relationOrderM}\\
g_\omega^{N}            & = & \left(T_1\right)^{a_{(\omega,1)}}\, \ldots\, \left(T_D\right)^{a_{(\omega,D)}}\;,\label{eqn:relationOrderN}\\
\com{g_\theta}{g_\omega}& = & \left(T_1\right)^{a_1}\,     \ldots\, \left(T_D\right)^{a_D}\;,\label{eqn:relationThetaOmega}\\
\com{T_i}{T_j}          & = & \Id_S \;,\label{eqn:relationTranslations}\\
\com{g_\theta}{T_i}     & = & \left(T_1\right)^{b_{(i,1)}}\, \ldots\, \left(T_D\right)^{b_{(i,D)}}\;,\label{eqn:RelationThetaTranslation}\\
\com{g_\omega}{T_i}     & = & \left(T_1\right)^{c_{(i,1)}}\, \ldots\, \left(T_D\right)^{c_{(i,D)}}\;,\label{eqn:RelationOmegaTranslation}
\end{eqnarray}
\end{subequations}
where $M$ and $N$ denote the order of $\theta$ and $\omega$, respectively, and 
$a_{(\theta,i)}, a_{(\omega,i)}, a_i, b_{(i,j)}, c_{(i,j)} \in \Z{}$ for $i,j = 1,\ldots, D$. A 
few remarks are in order: the right-hand sides of 
eqs.~\eqref{eqn:relationOrderM},~\eqref{eqn:relationOrderN} and~\eqref{eqn:relationThetaOmega} are 
non-trivial only in the case where $g_\theta = (\theta, \lambda_\theta)$ or 
$g_\omega = (\omega, \lambda_\omega)$ are roto-translations, i.e.\ $\lambda_\theta\not\in\Lambda_D$ 
or $\lambda_\omega\not\in\Lambda_D$. Furthermore, two translations $T_i=(\Id,e_i)$ and $T_j=(\Id,e_j)$ 
necessarily commute, see eq.~\eqref{eqn:relationTranslations}. Finally, the action of a twist (e.g.\ 
$\theta\in P$) on the lattice (e.g.\ $\hat\theta = e^{-1}\theta\, e$, see eq.~\eqref{eqn:TwistInLatticeBasis}) is 
uniquely specified by eqs.~\eqref{eqn:RelationThetaTranslation} and~\eqref{eqn:RelationOmegaTranslation}. 
In detail, comparing
\begin{equation}
\com{g_\theta}{T_i} ~=~ (\Id, (\Id - \theta^{-1})e_i)\;,
\end{equation}
with eq.~\eqref{eqn:RelationThetaTranslation} we obtain $\left(\hat\theta^{-1}\right)_{ij} = \delta_{ij}-b_{(j,i)}$.

Now, in order to write down a presentation of $S$, we rewrite each 
relation~\eqref{eqn:SpaceGroupRelations} such that one has the identity element $\Id_S$ on the 
right-hand side. For example, we modify eq.~\eqref{eqn:relationOrderM} to
\begin{equation}
g_\theta^{M}\, \left(T_1\right)^{-a_{(\theta,1)}}\, \ldots\, \left(T_D\right)^{-a_{(\theta,D)}} ~=~ \Id_S\;.
\end{equation}
Then, we suppress the identity element $\Id_S$ and do the same for all 
relations~\eqref{eqn:SpaceGroupRelations}. Consequently, a presentation of the space group $S$ reads
\begin{equation}\label{eqn:PresentationOfS}
S ~=~ \langle g_\theta,\, g_\omega,\, T_1,\, \ldots,\, T_D ~|~ g_\theta^{M}\, \left(T_1\right)^{-a_{(\theta,1)}}\, \ldots\, \left(T_D\right)^{-a_{(\theta,D)}},\, \ldots\,,\, \text{all relations} \rangle\;.
\end{equation}
One can use the presentation of a space group $S$ to uniquely specify $S$ without writing out the 
torus vielbein $e$ and the twist matrices explicitly. Indeed, the relations~\eqref{eqn:PresentationOfS} 
contain all information about the space group $S$. This fact is used in 
section~\ref{sec:spacegroupselectionrule} where we discuss the effective symmetries arising from the 
space group selection rule.

\subsection{Geometrical orbifolds and closed strings}
\label{app:StringsOnOrbifolds}

Having defined the space group $S$, a $D$-dimensional orbifold is defined geometrically as a quotient space 
$O = \mathbbm{R}^D/S$ using the equivalence relation
\begin{equation}\label{eqn:geometricalorbifold}
y_1 ~\sim~ y_2 \qquad \Leftrightarrow \qquad \exists~ g \in S \quad\text{such that}\quad y_1 = g\, y_2\;,
\end{equation}
for $y_1, y_2 \in \mathbbm{R}^D$. In words, two points $y_1$ and $y_2$ from $\mathbbm{R}^D$ are 
identified on the orbifold $O$ if there exists a space group element $g\in S$ that maps $y_2$ to 
$y_1$.

A closed string on an orbifold is characterized by the so-called constructing element 
$g=(\rho, \lambda)\in S$ that specifies the boundary condition for the string to close up to the 
action of $g$. For example, considering the worldsheet boson $X(\tau, \sigma)$ with worldsheet 
time and space coordinates $\tau$ and $\sigma \in [0,1]$, respectively, we impose the boundary 
condition
\begin{equation}\label{eqn:BoundaryCondition}
X(\tau,\sigma + 1) ~\stackrel{!}{=}~ g\, X(\tau,\sigma) ~=~ \rho\, X(\tau,\sigma) + \lambda\;.
\end{equation}
If $\rho\neq\Id$ the boundary condition 
eq.~\eqref{eqn:BoundaryCondition} describes a so-called twisted string that is localized at the 
fixed point of $g$. On the other hand, if $\rho=\Id$ the boundary condition 
eq.~\eqref{eqn:BoundaryCondition} describes a so-called untwisted string, which in general can be 
massless only if the winding vanishes, $\lambda=0$.

\begin{samepage}
Since $h\,X(\tau,\sigma)$ and $X(\tau,\sigma)$ are identified on the orbifold for all $h\in S$, 
see eq.~\eqref{eqn:geometricalorbifold}, the boundary condition eq.~\eqref{eqn:BoundaryCondition} 
with constructing element $g \in S$ and the corresponding one with constructing element 
$h\,g\,h^{-1} \in S$ describe the same string on the orbifold. Hence, a closed string with 
constructing element $g\in S$ is associated to the conjugacy class 
\begin{equation}
[g] ~=~ \{ h\,g\,h^{-1} ~|~ \text{ for all } h\in S\}\;.
\end{equation}
The resulting closed string state is denoted by $|[g]\rangle$.
\end{samepage}

Conventionally, one (ambiguously) labels an orbifold by the abstract finite group that is isomorphic 
to its point group $P$, for example $\Z{6}$ (and sometimes additional labels to distinguish between 
different representations of the same abstract finite group, for example $\Z{6}$-I or $\Z{6}$-II).
In terms of the modern nomenclature (see e.g.\ \cite{Fischer:2012qj}), this corresponds to the 
so-called $\mathbbm{Q}$-class. For a given $\mathbbm{Q}$-class, there can be several inequivalent 
torus lattices $\Lambda_D$, called $\mathbbm{Z}$-classes. Furthermore, for a given $\mathbbm{Z}$-class 
there can be several inequivalent roto-translations, called affine classes. Then, $\mathbbm{Z}$- 
and affine classes are consecutively enumerated. For example, the space group $\Z{2}\times\Z{2}$--2--5 
from table~\ref{tab:results} belongs to the $\mathbbm{Q}$-class $\Z{2}\times\Z{2}$, therein to the 
second $\mathbbm{Z}$-class and, finally, therein to the fifth affine class.

\acknowledgments

S.R-S. was partly supported by DGAPA-PAPIIT grant IN100217 and CONACyT grants F-252167
and 278017. P.V.\ is supported by the Deutsche Forschungsgemeinschaft (SFB1258).
S.R-S. and P.V. would like to thank Hans Peter Nilles for discussions and the Bethe Center for 
Theoretical Physics in Bonn for hospitality and support.


\begin{thebibliography}{10}

\bibitem{Dixon:1985jw}
L.~J. Dixon, J.~A. Harvey, C.~Vafa, and E.~Witten, \emph{{Strings on
  Orbifolds}}, Nucl. Phys. \textbf{B261} (1985), 678--686, [,678(1985)].

\bibitem{Dixon:1986jc}
L.~J. Dixon, J.~A. Harvey, C.~Vafa, and E.~Witten, \emph{{Strings on Orbifolds.
  2.}}, Nucl. Phys. \textbf{B274} (1986), 285--314.

\bibitem{Blaszczyk:2014qoa}
M.~Blaszczyk, S.~Groot~Nibbelink, O.~Loukas, and S.~Ramos-S{\'a}nchez,
  \emph{{Non-supersymmetric heterotic model building}}, JHEP \textbf{10}
  (2014), 119, \texttt{arXiv:1407.6362} [hep-th].

\bibitem{Blaszczyk:2015zta}
M.~Blaszczyk, S.~Groot~Nibbelink, O.~Loukas, and F.~Ruehle, \emph{{Calabi-Yau
  compactifications of non-supersymmetric heterotic string theory}}, JHEP
  \textbf{10} (2015), 166, \texttt{arXiv:1507.06147} [hep-th].

\bibitem{Lebedev:2006kn}
O.~Lebedev, H.~P. Nilles, S.~Raby, S.~Ramos-S{\'a}nchez, M.~Ratz, P.~K.~S.
  Vaudrevange, and A.~Wingerter, \emph{{A Mini-landscape of exact MSSM spectra
  in heterotic orbifolds}}, Phys. Lett. \textbf{B645} (2007), 88--94,
  \texttt{arXiv:hep-th/0611095} [hep-th].

\bibitem{Lebedev:2008un}
O.~Lebedev, H.~P. Nilles, S.~Ramos-S{\'a}nchez, M.~Ratz, and P.~K.~S.
  Vaudrevange, \emph{{Heterotic mini-landscape. (II). Completing the search for
  MSSM vacua in a $\Z{6}$ orbifold}}, Phys. Lett. \textbf{B668} (2008),
  331--335, \texttt{arXiv:0807.4384} [hep-th].

\bibitem{Pena:2012ki}
D.~K. Mayorga Pe\~na, H.~P. Nilles, and P.-K. Oehlmann, \emph{{A Zip-code for
  Quarks, Leptons and Higgs Bosons}}, JHEP \textbf{12} (2012), 024,
  \texttt{arXiv:1209.6041} [hep-th].

\bibitem{Nibbelink:2013lua}
S.~Groot~Nibbelink and O.~Loukas, \emph{{MSSM-like models on $\Z{8}$ toroidal
  orbifolds}}, JHEP \textbf{12} (2013), 044, \texttt{arXiv:1308.5145} [hep-th].

\bibitem{Nilles:2014owa}
H.~P. Nilles and P.~K.~S. Vaudrevange, \emph{{Geography of Fields in Extra
  Dimensions: String Theory Lessons for Particle Physics}}, Mod. Phys. Lett.
  \textbf{A30} (2015), no.~10, 1530008, \texttt{arXiv:1403.1597} [hep-th].

\bibitem{Carballo-Perez:2016ooy}
B.~Carballo-P{\'e}rez, E.~Peinado, and S.~Ramos-S{\'a}nchez,
  \emph{{$\Delta(54)$ flavor phenomenology and strings}}, JHEP \textbf{12}
  (2016), 131, \texttt{arXiv:1607.06812} [hep-ph].

\bibitem{Ramos-Sanchez:2017lmj}
S.~Ramos-S{\'a}nchez, \emph{{On flavor symmetries of phenomenologically viable
  string compactifications}}, J. Phys. Conf. Ser. \textbf{912} (2017), no.~1,
  012011, \texttt{arXiv:1708.01595} [hep-th].

\bibitem{Olguin-Trejo:2018wpw}
Y.~Olgu\'in-Trejo, R.~P\'erez-Mart\'inez, and S.~Ramos-S{\'a}nchez,
  \emph{{Charting the flavor landscape of MSSM-like Abelian heterotic
  orbifolds}},  (2018), \texttt{arXiv:1808.06622} [hep-th].

\bibitem{Nilles:2013lda}
H.~P. Nilles, S.~Ramos-S{\'a}nchez, M.~Ratz, and P.~K.~S. Vaudrevange, \emph{{A
  note on discrete $R$ symmetries in $\Z{6}$-II orbifolds with Wilson lines}},
  Phys. Lett. \textbf{B726} (2013), 876--881, \texttt{arXiv:1308.3435}
  [hep-th].

\bibitem{Bizet:2013wha}
N.~G. Cabo~Bizet, T.~Kobayashi, D.~K. Mayorga Pe\~na, S.~L. Parameswaran,
  M.~Schmitz, and I.~Zavala, \emph{{Discrete R-symmetries and Anomaly
  Universality in Heterotic Orbifolds}}, JHEP \textbf{02} (2014), 098,
  \texttt{arXiv:1308.5669} [hep-th].

\bibitem{Nilles:2017heg}
H.~P. Nilles, \emph{{Stringy Origin of Discrete R-symmetries}}, PoS
  \textbf{CORFU2016} (2017), 017, \texttt{arXiv:1705.01798} [hep-ph].

\bibitem{Lauer:1989ax}
J.~Lauer, J.~Mas, and H.~P. Nilles, \emph{{Duality and the Role of
  Nonperturbative Effects on the World Sheet}}, Phys. Lett. \textbf{B226}
  (1989), 251--256.

\bibitem{Lauer:1990tm}
J.~Lauer, J.~Mas, and H.~P. Nilles, \emph{{Twisted sector representations of
  discrete background symmetries for two-dimensional orbifolds}}, Nucl. Phys.
  \textbf{B351} (1991), 353--424.

\bibitem{Ibanez:1992hc}
L.~E. Ib{\'a\~n}ez and D.~L{\"u}st, \emph{{Duality anomaly cancellation,
  minimal string unification and the effective low-energy Lagrangian of 4-D
  strings}}, Nucl. Phys. \textbf{B382} (1992), 305--361,
  \texttt{arXiv:hep-th/9202046} [hep-th].

\bibitem{Bailin:1993ri}
D.~Bailin, A.~Love, W.~A. Sabra, and S.~Thomas, \emph{{Modular symmetries in
  Z(N) orbifold compactified string theories with Wilson lines}}, Mod. Phys.
  Lett. \textbf{A9} (1994), 1229--1238, \texttt{hep-th/9312122}.

\bibitem{Hamidi:1986vh}
S.~Hamidi and C.~Vafa, \emph{{Interactions on Orbifolds}}, Nucl. Phys.
  \textbf{B279} (1987), 465--513.

\bibitem{Dixon:1986qv}
L.~J. Dixon, D.~Friedan, E.~J. Martinec, and S.~H. Shenker, \emph{{The
  Conformal Field Theory of Orbifolds}}, Nucl. Phys. \textbf{B282} (1987),
  13--73.

\bibitem{Kobayashi:2006wq}
T.~Kobayashi, H.~P. Nilles, F.~Pl{\"o}ger, S.~Raby, and M.~Ratz, \emph{{Stringy
  origin of non-{A}belian discrete flavor symmetries}}, Nucl. Phys.
  \textbf{B768} (2007), 135--156, \texttt{arXiv:hep-ph/0611020} [hep-ph].

\bibitem{Nilles:2012cy}
H.~P. Nilles, M.~Ratz, and P.~K.~S. Vaudrevange, \emph{{Origin of Family
  Symmetries}}, Fortsch. Phys. \textbf{61} (2013), 493--506,
  \texttt{arXiv:1204.2206} [hep-ph].

\bibitem{Nilles:2018wex}
H.~P. Nilles, M.~Ratz, A.~Trautner, and P.~K.~S. Vaudrevange,
  \emph{{$\mathcal{CP}$ violation from string theory}}, Phys. Lett.
  \textbf{B786} (2018), 283--287, \texttt{arXiv:1808.07060} [hep-th].

\bibitem{Fischer:2012qj}
M.~Fischer, M.~Ratz, J.~Torrado, and P.~K.~S. Vaudrevange,
  \emph{{Classification of symmetric toroidal orbifolds}}, JHEP \textbf{01}
  (2013), 084, \texttt{arXiv:1209.3906} [hep-th], web page:
  http://users.ph.tum.de/ga57raj/Orbifolds/ClassificationOrbifolds/index.html.

\bibitem{Araki:2007zza}
T.~Araki, \emph{{Anomaly of Discrete Symmetries and Gauge Coupling
  Unification}}, Prog. Theor. Phys. \textbf{117} (2007), 1119--1138,
  \texttt{arXiv:hep-ph/0612306} [hep-ph].

\bibitem{Buchmuller:2006ik}
W.~Buchm{\"u}ller, K.~Hamaguchi, O.~Lebedev, and M.~Ratz, \emph{{Supersymmetric
  Standard Model from the Heterotic String (II)}}, Nucl. Phys. \textbf{B785}
  (2007), 149--209, \texttt{arXiv:hep-th/0606187} [hep-th].

\bibitem{Ratcliffe:2009}
J.~G. Ratcliffe and S.~T. Tschantz, \emph{Abelianization of space groups}, Acta
  Cryst. A \textbf{65} (2009), no.~1, 18--27,
  \texttt{https://onlinelibrary.wiley.com/doi/abs/10.1107/S0108767308036222}.

\bibitem{Blaszczyk:2012}
M.~Blaszczyk, \emph{{Heterotic Particle Models from various Perspectives}},
  Ph.D. thesis, University of Bonn, 2012,
  {http://hss.ulb.uni-bonn.de/2012/3021/3021.htm}.

\bibitem{Petersen:2009ip}
B.~Petersen, M.~Ratz, and R.~Schieren, \emph{{Patterns of remnant discrete
  symmetries}}, JHEP \textbf{08} (2009), 111, \texttt{arXiv:0907.4049}
  [hep-ph].

\bibitem{Donagi:2008xy}
R.~Donagi and K.~Wendland, \emph{{On orbifolds and free fermion
  constructions}}, J. Geom. Phys. \textbf{59} (2009), 942--968,
  \texttt{arXiv:0809.0330} [hep-th].

\bibitem{Forste:2006wq}
S.~F{\"o}rste, T.~Kobayashi, H.~Ohki, and K.-j. Takahashi,
  \emph{{Non-Factorisable $\Z{2}\times\Z{2}$ Heterotic Orbifold Models and
  {Y}ukawa Couplings}}, JHEP \textbf{03} (2007), 011,
  \texttt{arXiv:hep-th/0612044} [hep-th].

\bibitem{Blaszczyk:2009in}
M.~Blaszczyk, S.~Groot~Nibbelink, M.~Ratz, F.~Ruehle, M.~Trapletti, and
  P.~K.~S. Vaudrevange, \emph{{A $\Z{2}\times\Z{2}$ standard model}}, Phys.
  Lett. \textbf{B683} (2010), 340--348, \texttt{arXiv:0911.4905} [hep-th].

\bibitem{Nilles:2011aj}
H.~P. Nilles, S.~Ramos-S{\'a}nchez, P.~K.~S. Vaudrevange, and A.~Wingerter,
  \emph{{The Orbifolder: A Tool to study the Low Energy Effective Theory of
  Heterotic Orbifolds}}, Comput. Phys. Commun. \textbf{183} (2012), 1363--1380,
  \texttt{arXiv:1110.5229} [hep-th].

\bibitem{Fujikawa:1979ay}
K.~Fujikawa, \emph{{Path Integral Measure for Gauge Invariant Fermion
  Theories}}, Phys. Rev. Lett. \textbf{42} (1979), 1195--1198.

\bibitem{Fujikawa:1980eg}
K.~Fujikawa, \emph{{Path Integral for Gauge Theories with Fermions}}, Phys.
  Rev. \textbf{D21} (1980), 2848, [Erratum: Phys. Rev.D22,1499(1980)].

\bibitem{Araki:2008ek}
T.~Araki, T.~Kobayashi, J.~Kubo, S.~Ramos-S{\'a}nchez, M.~Ratz, and P.~K.~S.
  Vaudrevange, \emph{{(Non-)Abelian discrete anomalies}}, Nucl. Phys.
  \textbf{B805} (2008), 124--147, \texttt{arXiv:0805.0207} [hep-th].

\bibitem{Green:1984sg}
M.~B. Green and J.~H. Schwarz, \emph{{Anomaly Cancellation in Supersymmetric
  D=10 Gauge Theory and Superstring Theory}}, Phys. Lett. \textbf{149B} (1984),
  117--122.

\bibitem{Kakushadze:1996hj}
Z.~Kakushadze, G.~Shiu, and S.~H.~H. Tye, \emph{{Asymmetric nonAbelian
  orbifolds and model building}}, Phys. Rev. \textbf{D54} (1996), 7545--7560,
  \texttt{arXiv:hep-th/9607137} [hep-th].

\bibitem{Konopka:2012gy}
S.~J.~H. Konopka, \emph{{Non Abelian orbifold compactifications of the
  heterotic string}}, JHEP \textbf{07} (2013), 023, \texttt{arXiv:1210.5040}
  [hep-th].

\bibitem{Fischer:2013qza}
M.~Fischer, S.~Ramos-S{\'a}nchez, and P.~K.~S. Vaudrevange, \emph{{Heterotic
  non-Abelian orbifolds}}, JHEP \textbf{07} (2013), 080,
  \texttt{arXiv:1304.7742} [hep-th].

\end{thebibliography}

\providecommand{\bysame}{\leavevmode\hbox to3em{\hrulefill}\thinspace}

\end{document}